\documentclass[11pt]{article}
\usepackage{amsmath,amssymb}
\usepackage[dvipdfmx]{graphicx}
\usepackage{comment}

\begin{document}
\begin{titlepage}
\begin{flushright}
\end{flushright}
\begin{center}
  \vspace{3cm}
  {\bf \Large  Scalar Cosmological Perturbations in M-theory \\[0.2cm] with Higher Derivative Corrections}
  \\  \vspace{2cm}
  Kazuho Hiraga$^a$ and Yoshifumi Hyakutake$^b$
  \\ \vspace{1cm}
  {${}^a$\it Meijo University Senior High School,  \\
 Shintomi-cho 1-3-16, Nakamura-ku, 
Nagoya, Aichi 453-0031, Japan}\\
   \vspace{0.5cm}
  {${}^b$\it College of Science, Ibaraki University \\
   Bunkyo 2-1-1, Mito, Ibaraki 310-8512, Japan}
\end{center}

\vspace{2cm}
\begin{abstract}
We investigate the inflationary expansion of the universe induced by higher curvature corrections in M-theory.
The inflationary evolution of the geometry is discussed in ref.~\cite{Hiraga:2018kpb},
thus we succeed to analyse metric perturbations around the background.
Especially we focus on scalar perturbations and analyse linearized equations of motion for the scalar perturbations. 
By solving these equations explicitly, we evaluate the power spectrum of the curvature perturbation.
Scalar spectrum index is estimated under some assumption, and we show that it becomes close to 1.
\end{abstract}

\end{titlepage}

\setlength{\baselineskip}{0.65cm}


\section{Introduction} \label{sec:Intro}


Recent remarkable progress on astrophysical observations enables us to reveal the evolution of our universe\cite{Ade:2015lrj,Ade:2015tva,Array:2015xqh}.
Particularly these results support the inflationary scenario, in which the universe exponentially expands 
before the Big Bang\cite{Starobinsky:1980te,Guth:1980zm,Kazanas:1980tx,Sato:1980yn}.
There are a lot of models in which the inflation is caused by introducing a scalar field named inflaton field.
The inflaton field slowly rolls down its potential and the vacuum geometry becomes de-Sitter like\cite{Linde:1983gd}-\cite{Bezrukov:2007ep} 
(see also \cite{Kolb:1990vq}-\cite{Linde:2014nna} and references there).
Among a lot of inflationary models, superstring theory is a good candidate for the inflation scenario
since it possesses many scalar fields after the compactification.
Positions of D-branes are also described by scalar fields, and it is possible to identify one of these with the
inflation field\cite{Dvali:1998pa}-\cite{McAllister:2008hb} (see also \cite{Baumann:2014nda} and references there).

Besides the inflaton field, there are a lot of models in which the inflation is realized by modifying 
the gravity theory\cite{Starobinsky:1980te,Hwang:1996xh,DeFelice:2009ak,DeFelice:2010aj,Sebastiani:2016ras,Nojiri:2017ncd}.
Especially, the predictions of the Starobinsky model\cite{Starobinsky:1980te}, which contains curvature squared term in the action, 
are good agreement with the observations.
In fact, it predicts scalar spectral index $n_s=0.967$ and tensor to scalar ratio $r=0.003$ when the number of e-folds is 60. 
Since the curvature squared term in the Starobinsky model is considered as the quantum effect of the gravity,
it is natural to ask the origin of the quantum effect in more fundamental theory, such as the superstring theory or M-theory.
Actually heterotic superstring theory contains Gauss-Bonnet term, and type II superstring theories or M-theory contain quartic terms of 
the Riemann tensor\cite{Gross:1986iv}-\cite{Becker:2001pm}.
As examples, a study of the inflationary solutions in the heterotic superstring theory was done in ref.~\cite{Ishihara:1986if}, 
and studies of the inflationary solutions in the M-theory were executed in refs.~\cite{Ohta:2004wk,Maeda:2004vm,Maeda:2004hu,Akune:2006dg,Hiraga:2018kpb}.

In ref.~\cite{Hiraga:2018kpb}, we investigated the effect of leading curvature corrections in M-theory 
with respect to the homogeneous and isotropic geometry.
These higher curvature corrections are very important to evade no-go theorem for 
the existence of the de-Sitter like vacua in the string theory\cite{Gibbons:1984kp}-\cite{Obied:2018sgi}.
Actually we found that such corrections induce exponentially rapid expansion at early universe.
Furthermore the inflation naturally ends when the corrections are negligible compared to leading supergravity part.
Since the higher derivative corrections are universal in the superstring theory or the M-theory, the above is
a promising scenario to explain the origin of the inflation.
Therefore it is important to evaluate scalar spectral index and tensor to scalar ratio 
in the presence of the higher curvature corrections.
In this paper we propose a method to evaluate the scalar spectral index $n_s$ in the presence of higher derivative corrections.
And we show that $n_s$ is close to 1, if the power spectrum is constant at the beginning of the universe.

The organization of this paper is as follows.
In section \ref{sec:review}, we briefly review the inflationary scenario discussed in ref.~\cite{Hiraga:2018kpb}.
In section \ref{sec:ScalarPt}, first we consider perturbations around the background metric,
and examine their infinitesimal variations under the general coordinate transformation.
Next we derive linearized equations of motion for the perturbations.
Then we concentrate on the scalar perturbations and reduce their equations of motion by removing auxiliary fields.
In section \ref{sec:action}, we derive the second order effective action with respect to the scalar perturbations.
And we rederive the equations of motion for the scalar perturbations obtained in section \ref{sec:ScalarPt}.
In section \ref{sec:analyses}, we solve the equations of motion perturbatively and obtain explicit form of the 
curvature perturbation. Finally we evaluate the power spectrum of the curvature perturbation and 
calculate the scalar spectral index, which becomes close to 1.
Supplementary equations are listed in appendix \ref{sec:supp}\footnote{
Calculations in this paper are done by using Mathematica codes. See ref.~\cite{Mathematicacodes}},
dimensional reduction to 4 dimensional spacetime is explained in appendix \ref{sec:appB}, and thoughts on higher order terms are
discussed in appendix \ref{sec:appC}.


\section{Review of Inflationary Solution in M-theory} \label{sec:review}


In this section, we briefly review the inflationary solution in M-theory\cite{Hiraga:2018kpb}.
The effective action for the M-theory consists of supergravity part and higher derivative corrections.
Although the complete form of the higher derivative corrections is not known, we have control over leading curvature corrections.
Thus we truncate the effective action up to the leading higher curvature terms,
which is written as \cite{Tseytlin:2000sf,Becker:2001pm}
\begin{alignat}{3}
  S_{11} &= \frac{1}{2 \kappa_{11}^2} \int d^{11}x \; e \big( R + \Gamma Z \big), \label{eq:W4}
  \\[0.2cm]
  Z &\equiv 24 \big( W_{abcd} W^{abcd} W_{efgh} W^{efgh} - 64 W_{abcd} W^{aefg} W^{bcdh} W_{efgh} \notag
  \\[-0.1cm]
  &\qquad
  + 2 W_{abcd} W^{abef} W^{cdgh} W_{efgh} + 16 W_{acbd} W^{aebf} W^{cgdh} W_{egfh} \notag
  \\
  &\qquad
  - 16 W_{abcd} W^{aefg} W^b{}_{ef}{}^h W^{cd}{}_{gh} - 16 W_{abcd} W^{aefg} W^b{}_{fe}{}^h W^{cd}{}_{gh} \big), \notag
\end{alignat}
where $a,b,c, \cdots g,h$ are local Lorentz indices and $W_{abcd}$ is Weyl tensor.
There are two parameters, gravitational constant $2\kappa_{11}^2$ and expansion coefficient $\Gamma$, 
in the effective action. 
These are expressed in terms of 11 dimensional Planck length $\ell_p$ as
\begin{alignat}{3}
  2\kappa_{11}^2 = (2\pi)^8 \ell_p^9, \qquad \Gamma = \frac{\pi^2\ell_p^6}{2^{11} 3^2}. \label{eq:gamma} 
\end{alignat}
By varying the effective action (\ref{eq:W4}), we obtain following equations of motion\cite{Hyakutake:2013vwa}.
\begin{alignat}{3}
  E_{ab} &\equiv R_{ab} - \frac{1}{2} \eta_{ab} R 
  + \Gamma \Big\{ - \frac{1}{2} \eta_{ab} Z + R_{cdea} Y^{cde}{}_b - 2 D_{(c} D_{d)} Y^c{}_{ab}{}^d \Big\} = 0. \label{eq:MEOM}
\end{alignat}
Here $D_a$ is a covariant derivative with respect to the local Lorentz index.
The tensor $Y_{abcd}$ in the above is defined as
\begin{alignat}{3}
  Y_{abcd} &= X_{abcd} - \frac{1}{9} ( \eta_{ac} X_{bd} \!-\! \eta_{bc} X_{ad} \!-\! \eta_{ad} X_{bc} \!+\! \eta_{bd} X_{ac}) 
  + \frac{1}{90} ( \eta_{ac} \eta_{bd} \!-\! \eta_{ad} \eta_{bc} ) X, \label{eq:Ydef}
\end{alignat}
and the tensor $X_{abcd}$ is given by
\begin{alignat}{3}
  X_{abcd} &= \frac{1}{2} \big( X'_{[ab][cd]} + X'_{[cd][ab]} \big), \qquad 
  X_{ab} = X^c{}_{acb}, \qquad X = X^a{}_a, \label{eq:Xdef}
  \\
  X'_{abcd} &= 96 \big(
  W_{abcd} W_{efgh} W^{efgh} - 16 W_{abce} W_{dfgh} W^{efgh} + 2 W_{abef} W_{cdgh} W^{efgh} \notag
  \\
  &\qquad\,
  + 16 W_{aecg} W_{bfdh} W^{efgh} - 16 W_{abeg} W_{cf}{}^e{}_h W_d{}^{fgh} - 16 W_{efag} W^{ef}{}_{ch} W^g{}_b{}^h{}_d \notag
  \\
  &\qquad\,
  + 8 W_{ab}{}^{ef} W_{cegh} W_{df}{}^{gh} \big) \notag.
\end{alignat}
Note that $Y^c{}_{acb} = 0$.

Below we solve the equations of motion (\ref{eq:MEOM}) up to the linear order of $\Gamma$.
We assume that the 11 dimensional coordinates $X^\mu$ are divided into 4 dimensional spacetime $(t,x^i)$ and 7 internal directions $y^m$,
where $i=1,2,3$ and $m=4,\cdots,10$.
The ansatz for the metric is given by
\begin{alignat}{3}
  ds^2 &= - dt^2 + a(t)^2 dx_i^2 + b(t)^2 dy_m^2. \label{eq:bg}
\end{alignat}
$a(t)$ and $b(t)$ are scale factors of 3 dimensions and 7 internal ones, respectively\footnote{
The internal space is compactified on 7 dimensional torus with one volume factor $b$ and other moduli parameters are fixed 0.
Linear order perturbations of those are considered in section \ref{sec:ScalarPt}.
The dimensional reduction to 4 dimensional spacetime is explained in appendix \ref{sec:appB}.}.
Now we define Hubble parameter $H(t) = \frac{\dot{a}(t)}{a(t)}$ and similar one $G(t) = \frac{\dot{b}(t)}{b(t)}$. 
Then the equation (\ref{eq:MEOM}) is expressed in terms of $H(t)$ and $G(t)$, 
and the solution up to linear order of $\Gamma$ is given by
\begin{alignat}{3}
  H(\tau) &= \frac{H_\text{I}}{\tau} + \Gamma \frac{c_h H_\text{I}^7}{\tau^7}, \qquad
  G(\tau) &= \frac{-7 + \sqrt{21}}{14} \frac{H_\text{I}}{\tau} 
  + \Gamma \frac{c_g H_\text{I}^7}{\tau^7}. \label{eq:HGsol}
\end{alignat}
Here $\tau$ is dimensionless time coordinate given by
\begin{alignat}{3}
  \tau &= \frac{(-1 + \sqrt{21}) H_\text{I} t + 2}{2}, \label{eq:tau}
\end{alignat}
and numerical coefficients $c_h$ and $c_g$ are expressed as
\begin{alignat}{3}
  c_h &= \frac{13824 (477087 \!-\! 97732\sqrt{21})}{8575} \sim 47111, \label{eq:chcg}
  \\
  c_g &= - \frac{41472 (532196 \!-\! 110451 \sqrt{21})}{60025} \sim -17996. \notag
\end{alignat}
It is easy to integrate the eq.~(\ref{eq:HGsol}), and $\log a$ and $\log b$ are solved as
\begin{alignat}{3}
  \log a &= \log a_\text{E} + \frac{1+\sqrt{21}}{10} \log \tau 
  - \frac{1+\sqrt{21}}{60} c_h \Gamma H_\text{I}^6 \frac{1}{\tau^6}, \notag
  \\
  \log b &= \log b_\text{E} - \frac{3\sqrt{21}-7}{70} \log \tau - \frac{1+\sqrt{21}}{60} c_g \Gamma H_\text{I}^6 
  \frac{1}{\tau^6}. \label{eq:abQc}
\end{alignat}
From this we see that $a(\tau)$ is rapidly expanding and $b(\tau)$ is rapidly deflating during $1 \leq \tau \leq 2$.
$a_\text{E}$ or $b_\text{E}$ are integral constants and correspond to scale factors just after the inflation
or the deflation.
After $\tau=2$, the higher derivative corrections are suppressed and the scale factors behave like
\begin{alignat}{3}
  a_0 = a_\text{E} \, \tau^{\frac{1+\sqrt{21}}{10}}, \qquad
  b_0 = b_\text{E} \, \tau^{- \frac{3\sqrt{21}-7}{70}}. \label{eq:a0b0}
\end{alignat}
The behavior of $a_0$ is similar to radiation dominated era.

The motivation for the inflation is to resolve the horizon problem. This requires that the particle horizon $\int \frac{dt}{a(t)}$
during the inflationary era is almost equal to that after the radiation dominated era.
The particle horizon during the inflationary era is given by
\begin{alignat}{3}
  \frac{\sqrt{21}+1}{10 H_\text{I}} \int_1^2 \frac{d\tau}{a(\tau)} 
  &= \frac{\sqrt{21}+1}{10 a_\text{E} H_\text{I}} \int_1^2 d\tau \tau^{-\frac{1+\sqrt{21}}{10}}
  e^{\frac{1+\sqrt{21}}{60} c_h \Gamma H_\text{I}^6 \frac{1}{\tau^6}}. \label{eq:ph1}
\end{alignat}
On the other hand, if we simply apply the eq.~(\ref{eq:a0b0}) for the scale factor after $\tau=2$,
the particle horizon during this era is evaluated like
\begin{alignat}{3}
  \frac{\sqrt{21}+1}{10 H_\text{I}} \int_2^{\tau_0} \frac{d\tau}{a_0(\tau)} 
  &= \frac{\sqrt{21}+1}{10 a_\text{E} H_\text{I}} \int_2^{\tau_0} d\tau \tau^{-\frac{1+\sqrt{21}}{10}}
  \sim \frac{\sqrt{21}+1}{10 a_\text{E} H_\text{I}} \frac{9+\sqrt{21}}{6} \tau_0^{\frac{9-\sqrt{21}}{10}}, \label{eq:ph2}
\end{alignat}
where $\tau_0$ is the value at current time $t_0$.
Now we define the e-folding number as $N_\text{e} = \log \frac{a(\tau_0)}{a(2)}$.
This means that $\tau_0 = 2 e^{\frac{\sqrt{21}-1}{2} N_\text{e}}$. 
By equating the eq.~(\ref{eq:ph1}) with the eq.~(\ref{eq:ph2}), we obtain
\begin{alignat}{3}
  \int_1^2 d\tau \tau^{-\frac{1+\sqrt{21}}{10}} e^{\frac{1+\sqrt{21}}{60} c_h \Gamma H_\text{I}^6 \frac{1}{\tau^6}}
  &\sim \frac{9+\sqrt{21}}{6} 2^{\frac{9-\sqrt{21}}{10}} e^{\frac{\sqrt{21}-3}{2} N_\text{e}}. \label{eq:efnumber}
\end{alignat}
This gives a relation between $\Gamma H_\text{I}^6$ and $N_\text{e}$,
and we obtain $\Gamma H_\text{I}^6 \sim 0.014$ for $N_\text{e} = 69$, for example\footnote{Note that the definition
of the e-folding number is different from that in ref.~\cite{Hiraga:2018kpb}.}.


\section{Scalar Perturbations around the Background Geometry} \label{sec:ScalarPt}


Perturbations around homogeneous and isotropic universe are important directions to sort out inflationary models
via observations.
In this section, first we consider general metric perturbations around the background metric (\ref{eq:bg}),
and examine infinitesimal variations of perturbations under general coordinate transformation.
Next we derive linearized equations of motion for the perturbations.
Finally we focus on scalar perturbations and simplify their equations of motion
by removing auxiliary fields.
Main results are given by eqs.~(\ref{eq:P0}) and (\ref{eq:P1})\cite{Mathematicacodes}.

\subsection{Metric Perturbations and General Coordinate Transformation}

Let us consider perturbations around the background geometry (\ref{eq:bg}). 
We choose the metric with perturbations as follows.
\begin{alignat}{3}
  ds^2 &= - (1+2\alpha) dt^2 - 2 a(t) \beta_i dt dx^i + a(t)^2 (\delta_{ij} + h_{ij}) dx^i dx^j \notag
  \\
  &\quad\,
  - 2 b(t) \beta_m dt dy^m + b(t)^2 (\delta_{mn} + h_{mn}) dy^m dy^n + 2 a(t) b(t) h_{im} dx^i dy^m, \label{eq:metpt}
\end{alignat}
where $i,j = 1,2,3$ and $m,n=4,\cdots,10$.
$\alpha(X)$, $\beta_i(X)$ and $h_{ij}(X)$ are perturbations for the 4 dimensional spacetime,
and $h_{mn}(X)$ is that of the internal space. $\beta_m(X)$ and $h_{im}(X)$ are off-diagonal perturbations 
between 4 dimensional spacetime and the internal space.
As usual, we decompose vectors and tensors as
\begin{alignat}{3}
  \beta_i &= \hat{\beta}_i + \partial_i \beta, \qquad\;\;\;
  h_{ij} = \hat{h}_{ij} + 2 \partial_{(i} \hat{\gamma}_{j)} + 2 \partial_i \partial_j \gamma 
  + 2 \psi \delta_{ij},
  \\
  \beta_m &= \hat{\beta}_m + \partial_m \bar{\beta}, \qquad
  h_{mn} = \hat{h}_{mn} + 2 \partial_{(m} \hat{\gamma}_{n)} + 2 \partial_m \partial_n \bar{\gamma} 
  + 2 \bar{\psi} \delta_{mn}, \notag
  \\
  h_{im} &= \hat{h}_{im} + \partial_i \hat{\lambda}_m + \partial_m \hat{\lambda}_i + 2 \partial_i \partial_m \lambda. \notag
\end{alignat}
Here hatted vectors are divergenceless, and hatted tensors are divergenceless and traceless.
Note that $\hat{h}_{im}$ has $12$ independent components.
$\psi$ is the curvature perturbation whose power spectrum is important to sort out models via observations.

The general coordinate transformation is given by $X'^\mu = X^\mu + \xi^\mu(X)$. 
Again $\xi_i$ and $\xi_m$ are decomposed into
$\xi_i = \hat{\xi}_i + \partial_i \xi$ and $\xi_m = \hat{\xi}_m + \partial_m \bar{\xi}$, respectively.
Then the scalar perturbations transform as
\begin{alignat}{3}
  \beta' &= \beta - a^{-1} \xi^t + a \dot{\xi}, \qquad& 
  \bar{\beta}' &= \bar{\beta} - b^{-1} \xi^t + b \dot{\bar{\xi}}, \notag
  \\
  \gamma' &= \gamma - \xi, \qquad& 
  \bar{\gamma}' &= \bar{\gamma} - \bar{\xi}, \label{eq:gaugescalar}
  \\
  \psi' &= \psi - H \xi^t, \qquad& \bar{\psi}' &= \bar{\psi} - G \xi^t, \notag
  \\
  \alpha' &= \alpha-\dot{\xi}^t, \qquad& \lambda' &= \lambda - \tfrac{1}{2} a b^{-1} \xi - \tfrac{1}{2} a^{-1} b \bar{\xi}, \qquad & \notag
\end{alignat}
and vector perturbations do as
\begin{alignat}{3}
  \hat{\beta}'_i &= \hat{\beta}_i + a \dot{\hat{\xi}}_i, \qquad & \hat{\gamma}'_i &= \hat{\gamma}_i - \hat{\xi}_i, \qquad &
  \hat{\lambda}'_i &= \hat{\lambda}_i - a b^{-1} \hat{\xi}_i, \notag
  \\
  \hat{\beta}'_m &= \hat{\beta}_m + b \dot{\hat{\xi}}_m, \qquad & \hat{\gamma}'_m &= \hat{\gamma}_m - \hat{\xi}_m, \qquad &
  \hat{\lambda}'_m &= \hat{\lambda}_m - a^{-1} b \hat{\xi}_m. \label{eq:gaugevector}
\end{alignat}
Tensor perturbations do not transform under the general coordinate transformation.


\subsection{Equations of Motion for Metric Perturbations}


Let us derive linearized equations of motion for the metric perturbations.
This is simply done by varying the eq.~(\ref{eq:MEOM}).
First of all, variation of $Y_{abcd}$ is evaluated as
\begin{alignat}{3}
  \delta Y_{abcd} &= \delta X_{abcd} - \frac{1}{9} ( \eta_{ac} \delta X_{bd} \!-\! \eta_{bc} \delta X_{ad} 
  \!-\! \eta_{ad} \delta X_{bc} \!+\! \eta_{bd} \delta X_{ac}) 
  + \frac{1}{90} ( \eta_{ac} \eta_{bd} \!-\! \eta_{ad} \eta_{bc} ) \delta X, \label{eq:delYdef}
\end{alignat}
and variation of $X_{abcd}$ is given by
\begin{alignat}{3}
  \delta X_{abcd} &= \frac{1}{2} \big( \delta X'_{[ab][cd]} + \delta X'_{[cd][ab]} \big), \label{eq:delX}
  \\
  \delta X'_{abcd} 
  &= 96 \, \delta W_{pqrs} \big(
  \delta^p_a \delta^q_b \delta^r_c \delta^s_d W_{efgh} W^{efgh} 
  \!+\! 2 W_{abcd} W^{pqrs}
  \!-\! 16 \delta^p_a \delta^q_b \delta^r_c W_{dfgh} W^{sfgh}  \notag
  \\
  &\quad\,
  \!-\! 16 \delta^p_d W_{abce} W^{eqrs}
  \!-\! 16 W_{abc}{}^p W_d{}^{qrs}
  \!+\! 2 \delta^p_a \delta^q_b W_{cdgh} W^{rsgh}
  \!+\! 2 \delta^p_c \delta^q_d W_{abef} W^{efrs} \notag
  \\
  &\quad\,
  \!+\! 2 W_{ab}{}^{pq} W_{cd}{}^{rs}
  \!+\! 16 \delta^p_a \delta^r_c W_{bfdh} W^{qfsh}
  \!+\! 16 \delta^p_b \delta^r_d W_{aecg} W^{eqgs}
  \!+\! 16 W_a{}^p{}_c{}^r W_b{}^q{}_d{}^s \notag
  \\
  &\quad\,
  \!-\! 16 \delta^p_a \delta^q_b W_{cf}{}^r{}_h W_d{}^{fsh}
  \!-\! 16 \delta^p_c W_{ab}{}^r{}_g W_d{}^{qgs}
  \!-\! 16 \delta^p_d W_{abe}{}^r W_c{}^{qes}
  \!-\! 16 \delta^p_a W^{rs}{}_{ch} W^q{}_b{}^h{}_d \notag
  \\
  &\quad\,
  \!-\! 16 \delta^p_c W^{rs}{}_{ag} W^g{}_b{}^q{}_d
  \!-\! 16 \delta^p_b \delta^r_d W_{efa}{}^q W^{ef}{}_c{}^s 
  \!+\! 8 \delta^p_a \delta^q_b  W_c{}^r{}_{gh} W_d{}^{sgh}
  \!+\! 8 \delta^p_c W_{ab}{}^{qf} W_{df}{}^{rs} \notag
  \\
  &\quad\,
  \!+\! 8 \delta^p_d W_{ab}{}^{eq} W_{ce}{}^{rs} \big). \notag
\end{alignat}
Second, variation of $D_{(c} D_{d)} Y^c{}_{ab}{}^d$ is calculated as
\begin{alignat}{3}
  \delta \big( D_{c} D_{d} Y^c{}_{(ab)}{}^d \big)
  &= \delta e^\mu{}_c D_\mu D_{d} Y^c{}_{(ab)}{}^d
  + \delta \omega_{c}{}^c{}_e D_{d} Y^e{}_{(ab)}{}^d
  - \delta \omega_{ce(a} D^{d} Y^{ce}{}_{b)d} \notag
  \\
  &\quad\,
  - \delta \omega_{ce(a} D^{d} Y^c{}_{b)}{}^e{}_d 
  + D_c \delta \big( D_d Y^c{}_{(ab)}{}^d \big) \notag
  \\[0.1cm]
  &\!= \delta e^\mu{}_c D_\mu D_{d} Y^c{}_{(ab)}{}^d
  \!+\! \delta \omega_{c}{}^c{}_e D_{d} Y^e{}_{(ab)}{}^d 
  \!-\! \delta \omega_{ce(a} D^{d} Y^{ce}{}_{b)d} \notag
  \\
  &\quad\,
  - \delta \omega_{ce(a} D^{d} Y^c{}_{b)}{}^e{}_d 
  + D_c D_d \delta Y^c{}_{(ab)}{}^d
  + D_c \big( \delta e^\nu{}_d D_\nu Y^c{}_{(ab)}{}^d \label{eq:delDDY}
  \\
  &\quad\,
  + \delta \omega_d{}^c{}_e Y^e{}_{(ab)}{}^d
  - \delta \omega_{de(a} Y^{ce}{}_{b)}{}^d 
  - \delta \omega_{de(a} Y^c{}_{b)}{}^{ed} 
  + \delta \omega_d{}^d{}_e Y^c{}_{(ab)}{}^e \big), \notag 
\end{alignat}
where $\delta \omega_a{}^b{}_c \equiv e^\mu{}_a \delta \omega_\mu {}^b{}_c$. 
Combining the above results, we see that the variation of the eq.~(\ref{eq:MEOM}) is evaluated as
\begin{alignat}{3}
  \delta E_{ab} &= \delta R_{ab} \!-\! \frac{1}{2} \eta_{ab} \delta R 
  + \Gamma \Big\{ \!\!-\! \frac{1}{2} \eta_{ab} \delta R_{cdef} Y^{cdef}
  \!+\! \delta R_{cdea} Y^{cde}{}_b \!+\! R^{cde}{}_a \delta Y_{cdeb} \notag
  \\
  &\quad\,
  - 2 \delta e^\mu{}_c D_\mu D_{d} Y^c{}_{(ab)}{}^d
  - 2 \delta \omega_{c}{}^c{}_e D_{d} Y^e{}_{(ab)}{}^d 
  + 2 \delta \omega_{ce(a} D^{d} Y^{ce}{}_{b)d} \notag
  \\
  &\quad\,
  + 2 \delta \omega_{ce(a} D^{d} Y^c{}_{b)}{}^e{}_d 
  - 2 D_c D_d \delta Y^c{}_{(ab)}{}^d 
  \!-\! 2 D_c \big( \delta e^\nu{}_d D_\nu Y^c{}_{(ab)}{}^d \label{eq:MEOMvar}
  \\
  &\quad\,
  + \delta \omega_d{}^c{}_e Y^e{}_{(ab)}{}^d
  - \delta \omega_{de(a} Y^{ce}{}_{b)}{}^d 
  - \delta \omega_{de(a} Y^c{}_{b)}{}^{ed} 
  + \delta \omega_d{}^d{}_e Y^c{}_{(ab)}{}^e \big) 
  \Big\} = 0. \notag
\end{alignat}


\subsection{Equations of Motion for Scalar Perturbations}


In this subsection, we restrict the metric perturbations to the scalar perturbations. 
So the metric is chosen as
\begin{alignat}{3}
  ds^2 &= - (1+2\alpha) dt^2 - 2 a \partial_i \beta dt dx^i 
  + a^2 (\delta_{ij} + 2 \partial_i \partial_j \gamma + 2 \psi \delta_{ij}) dx^i dx^j \notag
  \\
  &\quad\,
  - 2 b \partial_m \bar{\beta} dt dy^m 
  + b^2 (\delta_{mn} + 2 \partial_m \partial_n \bar{\gamma} + 2 \bar{\psi} \delta_{mn}) dy^m dy^n 
  + 4 a b \partial_i \partial_m \lambda dx^i dy^m. \label{eq:scalarpt}
\end{alignat}
If we choose some gauge, the above metric is equivalent to the vielbein of the form
\begin{alignat}{3}
  &e^a{}_\mu + \delta e^a{}_\mu \label{eq:vbPt}
  \\
  &\!=\! \begin{pmatrix} 
    1 \!+\! \alpha & 0 & 0 & 0 & 0 & \cdots & 0 \\
    \!- \partial_1 \beta & \!\!\!a(1 \!+\! \partial_1^2 \gamma \!+\! \psi)\!\!\! & 0 & 0 & 0 & \cdots & 0 \\
    \!- \partial_2 \beta & 2 a \partial_2 \partial_1 \gamma & \!\!\!a(1 \!+\!  \partial_2^2 \gamma \!+\!  \psi)\!\!\! & 0 & 0 & \cdots & 0 \\
    \!- \partial_3 \beta & 2 a \partial_3 \partial_1 \gamma & 2 a \partial_3 \partial_2 \gamma & \!\!\!a(1 \!+\! \partial_3^2 \gamma \!+\! \psi)\!\!\! & 0 & \cdots & 0 \\
    \!- \partial_4 \bar{\beta} & 2 a  \partial_4 \partial_1 \lambda & 2 a  \partial_4 \partial_2 \lambda & 2 a \partial_4 \partial_3 \lambda & \!\!\!b(1 \!+\! \partial_4^2 \bar{\gamma} \!+\! \bar{\psi})\!\!\! & \cdots & 0 \\
    \vdots & \vdots & \vdots & \vdots & \vdots & \ddots & 0 \\
    \!- \partial_{10} \bar{\beta} & 2 a \partial_{10} \partial_1 \lambda & 2 a \partial_{10} \partial_2 \lambda & 2 a \partial_{10} \partial_3 \lambda & 2 b \partial_{10} \partial_4 \bar{\gamma} & \cdots & \!\!\!b(1 \!+\! \partial_{10}^2 \bar{\gamma} \!+\! \bar{\psi})\! \notag
  \end{pmatrix},
\end{alignat}
up to the linear order of the perturbations. 
Here $e^a{}_\mu$ is the background vielbein and $\delta e^a{}_\mu$ linearly depends on the scalar perturbations.

Now we define following quantities.
\begin{alignat}{3}
  \chi &= a (\beta + a \dot{\gamma}),& \qquad \bar{\chi} &= b (\bar{\beta} + b \dot{\bar{\gamma}}), \label{eq:chisigma}
  \\
  \Psi &= H^{-1}\psi, & \qquad \bar{\Psi} &= G^{-1} \bar{\psi}, \qquad\qquad
  \sigma = ab \lambda - \frac{a^2}{2} \gamma - \frac{b^2}{2} \bar{\gamma}. \notag
\end{alignat}
Consulting the eq.~(\ref{eq:gaugescalar}), we see that $\alpha, \chi, \Psi, \bar{\chi}, \bar{\Psi}$ are invariant under $\xi$ and $\bar{\xi}$ transformations,
and $\sigma$ is invariant under the general coordinate transformation. 
By inserting scalar perturbations (\ref{eq:vbPt}) into eq.~(\ref{eq:MEOMvar}), 
and expanding all perturbations by Fourier modes, such as
\begin{alignat}{3}
  \Psi(t,x,y) &= \int d^3k d^7l \, \big\{ \Psi(t,k,l) e^{i k_i x^i + i l_m y^m} +
  \Psi(t,k,l)^\ast e^{-i k_i x^i - i l_m y^m} \big\}, \label{eq:FmPsi}
\end{alignat}
we obtain following 8 linearized equations with respect to
$\tilde{\Upsilon} = \{\alpha(t,k,l)$, $\chi(t,k,l)$, $\Psi(t,k,l)$, $\bar{\chi}(t,k,l)$, $\bar{\Psi}(t,k,l)$, $\sigma(t,k,l)\}$. 
\begin{alignat}{3}
  E_1 &\!\equiv\! \delta E_{00} \!=\! 0, \quad 
  E_2 \!\equiv\! \frac{a}{k_a} \delta E_{0a} \!=\! 0, \quad
  E_3 \!\equiv\! \frac{a^2}{k_a k_b} \delta E_{ab} \!=\! 0, \quad 
  E_4 \!\equiv\! \delta E_{aa} - \frac{k_a}{k_b} \delta E_{ab} \!=\! 0, \label{eq:8EOMs}
  \\
  E_5 &\!\equiv\! \frac{b}{l_m} \delta E_{0m} \!=\! 0, \quad
  E_6 \!\equiv\! \frac{b^2}{l_m l_n} \delta E_{mn} \!=\! 0, \quad
  E_7 \!\equiv\! \delta E_{mm} - \frac{l_m}{l_n} \delta E_{mn} \!=\! 0, \quad
  E_8 \!\equiv\! \frac{ab}{k_a l_m} \delta E_{am} \!=\! 0. \notag
\end{alignat}
Here the indices are not contracted. 
In order to evaluate the above equations we employed Mathematica codes,
and portions of results are written as
\begin{alignat}{3}
  E_1 &= \frac{k^2}{a^2} \Big\{ - (7 G + 2 H) \chi + 2 H \Psi + 7 G \bar{\Psi } \Big\}
  + \frac{l^2}{b^2} \Big\{ - 3 (2G + H) \bar{\chi } + 3 H \Psi + 6 G \bar{\Psi} \Big\} \notag
  \\
  &\quad\,
  + \frac{k^2 l^2 }{a^2 b^2} 2 \sigma
  - 6 (7 G^2+7 G H+H^2) \alpha + 3 \dot{H} (7 G+2 H) \Psi + 3 H (7 G+2 H) \dot{\Psi } \label{eq:E1}
  \\
  &\quad\,
  + 21 \dot{G} (2 G+H) \bar{\Psi } + 21 G (2G+H) \dot{\bar{\Psi}} + \Gamma \tilde{S}_1(\tilde{\Upsilon},H,G), \notag
  \\[0.2cm]
  E_2 &= \frac{l^2}{b^2} \Big\{ -\frac{\chi}{2} + \frac{\bar{\chi}}{2} + 2 H \sigma -\dot{\sigma} \Big\}
  + (7 G+2 H) \alpha - 2 \dot{H} \Psi - 2 H \dot{\Psi} \label{eq:E2}
  \\
  &\quad\,
  - 7 (G^2-G H+\dot{G}) \bar{\Psi} - 7 G \dot{\bar{\Psi}} + \Gamma \tilde{S}_2(\tilde{\Upsilon},H,G), \notag
  \\[0.2cm]
  E_3 &= - \frac{2 l^2 \sigma}{b^2} - \alpha + (7 G+H) \chi + \dot{\chi} - H \Psi - 7 G \bar{\Psi} 
  + \Gamma \tilde{S}_3(\tilde{\Upsilon},H,G), \label{eq:E3}
\end{alignat}
\vspace{-1cm}
\begin{alignat}{3}
  E_4 &= \frac{k^2}{a^2} \Big\{ - \alpha + (7 G+H) \chi + \dot{\chi} - H \Psi - 7 G \bar{\Psi} \Big\} \notag
  \\
  &\quad\,
  + \frac{l^2}{b^2} \Big\{ - \alpha + 2(3 G+H) \bar{\chi} + \dot{\bar{\chi}} - 2 H \Psi - 6 G \bar{\Psi} \Big\} 
  - \frac{k^2 l^2 }{a^2 b^2} 2 \sigma \notag
  \\
  &\quad\,
  + 2 (28 G^2 + 14 G H + 3 H^2 + 7 \dot{G} + 2\dot{H}) \alpha + (7 G+2 H) \dot{\alpha} \notag
  \\
  &\quad\,
  - 2 (7 G\dot{H}+\ddot{H}+3 H \dot{H}) \Psi - 2 (7 G H+3 H^2+2 \dot{H}) \dot{\Psi} - 2 H \ddot{\Psi } 
  \label{eq:E4}
  \\
  &\quad\,
  - 7 (2 \dot{G} H+\ddot{G}+8 G \dot{G}) \bar{\Psi } - 14 (4 G^2+G H+\dot{G}) \dot{\bar{\Psi}} 
  - 7 G \ddot{\bar{\Psi}} 
  + \Gamma \tilde{S}_4(\tilde{\Upsilon},H,G), \notag
  \\[0.2cm]
  E_5 &= \frac{k^2}{a^2} \Big\{ \frac{\chi}{2} - \frac{\bar{\chi}}{2} + 2 G \sigma - \dot{\sigma} \Big\}
  + 3 (2 G+H) \alpha + 3 (G H-H^2-\dot{H}) \Psi - 3 H \dot{\Psi} \qquad\, \notag
  \\
  &\quad\,
  - 6 \dot{G} \bar{\Psi} - 6 G \dot{\bar{\Psi}} + \Gamma \tilde{S}_5(\tilde{\Upsilon},H,G), \label{eq:E5}
  \\[0.2cm]
  E_6 &= - \frac{k^2}{a^2} 2 \sigma - \alpha + (5 G+3 H) \bar{\chi} + \dot{\bar{\chi}} 
  - 3 H \Psi - 5 G \bar{\Psi} + \Gamma \tilde{S}_6(\tilde{\Upsilon},H,G), \label{eq:E6}
  \\[0.2cm]
  E_7 &= \frac{k^2}{a^2} \Big\{ - \alpha + (6 G+2 H) \chi + \dot{\chi} - 2 H \Psi - 6 G \bar{\Psi} \Big\} \notag
  \\
  &\quad\,
  + \frac{l^2}{b^2} \Big\{ - \alpha + (5 G+3 H) \bar{\chi} + \dot{\bar{\chi}} - 3 H \Psi - 5 G \bar{\Psi} \Big\} 
  - \frac{k^2 l^2}{a^2 b^2} 2 \sigma \notag
  \\
  &\quad\,
  + 6 \big( 7 G^2 + 6 G H + 2 H^2 + 2 \dot{G} + \dot{H} \big) \alpha 
  + 3 (2 G+H) \dot{\alpha} \notag
  \\
  &\quad\,
  - 3 (6 G \dot{H}+\ddot{H}+4 H \dot{H}) \Psi - 6 (3 G H+2 H^2+\dot{H}) \dot{\Psi} - 3 H \ddot{\Psi } \label{eq:E7}
  \\
  &\quad\,
  - 6 (3 \dot{G} H+\ddot{G}+7 G\dot{G}) \bar{\Psi} - 6 (7 G^2+3 G H+2 \dot{G}) \dot{\bar{\Psi}} - 6 G \ddot{\bar{\Psi}} 
  + \Gamma \tilde{S}_7(\tilde{\Upsilon},H,G), \quad \notag
  \\[0.2cm]
  E_8 &= - \alpha + \frac{1}{2} (5 G+3 H) \chi + \frac{1}{2} \dot{\chi} + \frac{1}{2} (7 G+H) \bar{\chi}  
  + \frac{1}{2} \dot{\bar{\chi}} - 2 H \Psi - 6 G \bar{\Psi} \notag
  \\
  &\quad\,
  - 2 (7 G^2+G H+\dot{G}) \sigma + (5G+H) \dot{\sigma} + \ddot{\sigma} + \Gamma \tilde{S}_8(\tilde{\Upsilon},H,G), \label{eq:E8}
\end{alignat}
where $k^2 = k_i k^i$ and $l^2 = l_m l^m$.
Note that one of these equations is redundant because of $D^a E_{ab} = 0$, so we have 7 independent equations.
The explicit forms of $\tilde{S}_u (u=1,\cdots,8)$ can be found in ref.~\cite{Mathematicacodes}.

Now we simply set $l_m = 0$, because $\frac{l^2}{b^2}$ becomes very large after the rapid expansion
and such massive modes will decouple.
Anyway the full analyses including $l_i \neq 0$ modes will be discussed elsewhere.
When $l_i = 0$, the above 7 independent equations reduce to following 4 equations, 
which are linear on $\Upsilon = \{\alpha,\chi,\Psi,\bar{\Psi}\}$.
\begin{alignat}{3}
  E_1 &= \frac{k^2}{a^2} \Big\{ - (7 G+2 H) \chi + 2 H \Psi + 7 G \bar{\Psi} \Big\} \notag
  \\
  &\quad\,
  - 6 (7 G^2+7 G H+H^2) \alpha + 3 \dot{H} (7 G+2 H) \Psi + 3 H (7 G+2 H) \dot{\Psi} \label{eq:E12}
  \\
  &\quad\,
  + 21 \dot{G} (2 G+H) \bar{\Psi} + 21 G (2G+H) \dot{\bar{\Psi}} + \Gamma S_1(\Upsilon,H,G) = 0, \notag
  \\[0.2cm]
  E_2 &= (7 G+2 H) \alpha - 2 \dot{H} \Psi - 2 H \dot{\Psi} 
  - 7 (G^2-G H+\dot{G}) \bar{\Psi} - 7 G \dot{\bar{\Psi}} + \Gamma S_2(\Upsilon,H,G) = 0, \label{eq:E22}
  \\[0.2cm]
  E_3 &= - \alpha + (7 G+H) \chi + \dot{\chi} - H \Psi - 7 G \bar{\Psi} + \Gamma S_3(\Upsilon,H,G) = 0, 
  \label{eq:E32}
\end{alignat}
\begin{alignat}{3}
  E_7 &= \frac{k^2}{a^2} \Big\{ - \alpha + (6 G+2 H) \chi + \dot{\chi} - 2 H \Psi - 6 G \bar{\Psi} \Big\} \notag
  \\
  &\quad\,
  + 6 ( 7 G^2 + 6 G H + 2 H^2 + \dot{G} + \dot{H} ) \alpha + 3 (2 G+H) \dot{\alpha} \label{eq:E72}
  \notag
  \\
  &\quad\,
  - 3 (6 G \dot{H}+\ddot{H}+4 H \dot{H}) \Psi - 6 (3 G H+2 H^2+\dot{H}) \dot{\Psi} - 3 H \ddot{\Psi}
  \\
  &\quad\,
  - 6 (3 \dot{G} H+\ddot{G}+7 G\dot{G}) \bar{\Psi} - 6 (7 G^2+3 G H+2 \dot{G}) \dot{\bar{\Psi}} - 6 G \ddot{\bar{\Psi}} 
  + \Gamma S_7(\Upsilon,H,G)  = 0. \notag
\end{alignat}
Again the explicit forms of $S_u (u=1,2,3,7)$ can be found in ref.~\cite{Mathematicacodes}.

Let us solve above equations up to the linear order of $\Gamma$ expansion, 
so we expand the perturbation $\Upsilon$ as $\Upsilon = \Upsilon_0 + \Gamma \Upsilon_1$.
Here the subscript $0$ represents the quantity at the leading order, and $1$ does one which is linear order of $\Gamma$.
As will be clear soon, it is useful to define following gauge invariant quantity.
\begin{alignat}{3}
  P \equiv \Psi - \bar{\Psi} = H^{-1} \psi - G^{-1} \bar{\psi}. \label{eq:defP}
\end{alignat}
Below we will show that equations of motion for scalar perturbations can be collected into
single differential equation with respect to $P$, after eliminating auxiliary fields $\alpha$ and $\chi$.

First let us consider the equations of motion at the leading order of $\Gamma$ expansion. 
From eqs.~(\ref{eq:E12}) and (\ref{eq:E22}), $\alpha_0$ and $\chi_0$ are solved as
\begin{alignat}{3}
  \alpha_0 &= - \frac{9+\sqrt{21}}{3} H_0 P_0 + \dot{\Psi}_0  + \frac{\sqrt{21}}{3} \dot{P}_0 , \label{eq:alpha0}
  \\
  \chi_0 &= \frac{a^2}{k^2} \Big\{ - \frac{3(\sqrt{21}-1)}{2} H_0^2 P_0 + 3 H_0 \dot{P}_0 \Big\} 
  + \Psi_0 + \frac{\sqrt{21}}{3} P_0. \label{eq:chi0}
\end{alignat}
Here $H_0$ and $G_0$ are leading parts of $H$ and $G$, respectively, and we used 
$G_0 = \frac{-7 + \sqrt{21}}{14} H_0$ and $\dot{H}_0 = \frac{1 - \sqrt{21}}{2} H_0^2$.
By inserting the above into the eq.~(\ref{eq:E32}) with $\Gamma=0$, we obtain 
\begin{alignat}{3}
  0 &= \ddot{P}_0 - \frac{\sqrt{21}-1}{2} H_0 \dot{P}_0 
  + \Big( \frac{k^2}{a_0^2} - \frac{\sqrt{21}-11}{2} H_0^2 \Big) P_0. \label{eq:P0}
\end{alignat}
Note that the eq.~(\ref{eq:E72}) is automatically satisfied. 
So we only need to solve the eq.~(\ref{eq:P0}) at the leading order of $\Gamma$.

Next let us investigate linear order of $\Gamma$ expansion. 
Again, from eqs.~(\ref{eq:E12}) and (\ref{eq:E22}), auxiliary fields $\alpha$ and $\chi$ are 
solved up to linear order of $\Gamma$ as
\begin{alignat}{3}
  \alpha &= 
  \frac{2 \dot{H} \Psi + 2 H \dot{\Psi} + 7( \dot{G} - G H + G^2) \bar{\Psi} + 7 G \dot{\bar{\Psi}}}{7 G+2 H} \notag
  \\
  &\quad\,
  + \Gamma \Big[ \tfrac{1536 (14229047+734623 \sqrt{21}) }{8575} H_0^7 \Psi_0
  + \tfrac{3072 (828991\sqrt{21}-14799601) }{8575} H_0^7 \bar{\Psi}_0 \notag
  \\
  &\quad\,
  + \tfrac{1536 (-2965613-1496367 \sqrt{21} ) }{8575} H_0^6 \dot{P}_0
  + \tfrac{546816 (136+9 \sqrt{21}) }{245} H_0^5 \ddot{P}_0 
  - \tfrac{6144 (1771+1014\sqrt{21}) }{1225} H_0^4 \dddot{P}_0 \notag
  \\
  &\quad\,
  - \frac{k^2}{a^2} \Big\{ \tfrac{1536 (1677 \sqrt{21}-152147) }{1225} H_0^5 P_0 
  + \tfrac{3072 (497+1348 \sqrt{21}) }{1225} H_0^4 \dot{P}_0 \Big\} \Big], \label{eq:alpha}
\end{alignat}
\begin{alignat}{3}
  \chi &= 
  \frac{a^2}{k^2} \frac{ 63 G^2 \dot{H} \Psi - 21 G ( 3 H \dot{G} \!-\! 12 G H^2 \!+\! 14 G^3 \!-\! 2 H^3 ) \bar{\Psi}
  + 63 G^2 H \dot{P} }{(7 G + 2 H)^2}
  + \frac{2 H \Psi + 7 G \bar{\Psi}}{7 G + 2 H} \notag
  \\
  &\quad
  + \Gamma \Big[  
  \frac{a^2}{k^2} \Big\{ - \tfrac{3072 (34956397-13586977 \sqrt{21}) }{8575} H_0^8 \Psi_0 
  - \tfrac{3072 (26812583+1603297 \sqrt{21}) }{8575} H_0^8 \bar{\Psi}_0 \notag
  \\
  &\quad\,
  - \tfrac{3072 (25037012-2136942 \sqrt{21}) }{8575} H_0^7 \dot{P}_0 
  - \tfrac{3072 (628026-719166 \sqrt{21}) }{8575} H_0^6 \ddot{P}_0 \notag
  \\
  &\quad\,
  - \tfrac{3072 (34463+217\sqrt{21}) }{1225} H_0^5 \dddot{P}_0 \Big\} 
  + \tfrac{3072 (5902\sqrt{21}-573447) }{8575} H_0^6 P_0 \notag
  \\
  &\quad\,
  + \tfrac{4608 (861+859 \sqrt{21}) }{245} H_0^5 \dot{P}_0
  - \tfrac{6144 (1771+1014\sqrt{21}) }{1225} H_0^4 \ddot{P}_0 
  - \frac{k^2}{a^2} \tfrac{3072 (497+1348 \sqrt{21}) }{1225} H_0^4 P_0 \Big]. \label{eq:chi}
\end{alignat}
By inserting the above into (\ref{eq:E32}), of course we obtain the eq.~(\ref{eq:P0}) at the leading order, and
\begin{alignat}{3}
  0 &= \ddot{P}_1 - \frac{\sqrt{21}-1}{2} H_0 \dot{P}_1 
  + \Big( \frac{k^2}{a_0^2} - \frac{\sqrt{21}-11}{2} H_0^2 \Big) P_1 \notag
  \\
  &\quad\,
  - \tfrac{1536 (49692383 \sqrt{21}-70593438) }{8575} H_0^8 P_0
  + \tfrac{768 (36412229 \sqrt{21}-124991079) }{1715} H_0^7 \dot{P}_0 \notag
  \\
  &\quad\,
  + \tfrac{768 (5604373\sqrt{21}-36068337) }{1715} H_0^6 \ddot{P}_0
  + \tfrac{12288 (6383 \sqrt{21}-17688) }{245} H_0^5 \dddot{P}_0 \notag
  \\
  &\quad\,
  + \tfrac{3072 (2261\sqrt{21}-23271) }{1225} H_0^4 \ddddot{P}_0
  + \frac{k^2}{a_0^2} \Big\{ \Big( \tfrac{768 (3567079\sqrt{21}-29260239) }{8575} H_0^6 
  - 2 \bar{a}_1 \Big) P_0 \notag
  \\
  &\quad\,
  + \tfrac{3072 (85331 \sqrt{21}-265416) }{1225} H_0^5 \dot{P}_0
  + \tfrac{1536 (9479 \sqrt{21}-66369) }{1225} H_0^4 \ddot{P}_0 \Big\} \label{eq:P1}
  \\
  &\quad\,
  + \frac{k^4}{a_0^4} \tfrac{6144 (1633 \sqrt{21}-9288) }{1225} H_0^4 P_0, \notag
\end{alignat}
at the linear order of $\Gamma$.
Here $\bar{a}_1$, which comes from $\frac{k^2}{a^2} P_0$, is defined as
\begin{alignat}{3}
  \bar{a}_1 = - \frac{1+\sqrt{21}}{60} c_h \frac{H_\text{I}^6}{\tau^6}
  = - \frac{1+\sqrt{21}}{60} c_h H_0^6. \label{a1bar}
\end{alignat}
In summary, we have derived the eq.~(\ref{eq:P0}) and the eq.~(\ref{eq:P1}) for the scalar perturbation $P$.
We will solve these equations up to the linear order of $\Gamma$ in section \ref{sec:analyses}.


\section{Effective Action for Scalar Perturbations} \label{sec:action}


In the previous section, we derived equations of motion for scalar perturbations, which are expressed by the eq.~(\ref{eq:P0}) and the eq.~(\ref{eq:P1}).
In this section, we consider effective action which is second order with respect to the scalar perturbations. 
We will reproduce the equations of motion (\ref{eq:P0}) and eq.~(\ref{eq:P1}) from this effective action.

First let us substitute the metric (\ref{eq:metpt}) into the action (\ref{eq:W4}), and expand it up to the second order with respect to 
the scalar perturbations.
By setting $l_m=0$, the result is written as
\begin{alignat}{3}
  S^{(2)}_\text{pt} &= S^{(2,0)}_\text{pt} + \Gamma S^{(2,1)}_\text{pt} \notag
  \\
  &= \frac{1}{2 \kappa_{11}^2} \int dt d^3x d^7y \, a^3 b^7 \Big[ 
  - 42 G^2 \dot{\bar{\Psi }}^2 - 6 H^2 \dot{\Psi }^2 - 42  G H \dot{\Psi } \dot{\bar{\Psi }} \notag
  \\
  &\quad\,
  + 42 \dot{\Psi } \bar{\Psi }  \big( - G^2 H + G H^2 + G \dot{H} - \dot{G} H \big) \notag
  \\
  &\quad\,
  + 42 G \bar{\Psi} ^2 \big(21 G^3 + 18 G^2 H + 6 G H^2 + 3 G \dot{H} + 13 \dot{G} G + 3 \dot{G} H + \ddot{G} \big) \notag
  \\
  &\quad\,
  + 42 G \Psi \bar{\Psi} \big(21 G^2 H + 18 G H^2 + 6 G \dot{H} + 6 H^3 + \ddot{H} + 6 \dot{G} H + 7 H \dot{H} \big) \notag
  \\
  &\quad\, 
  + 6 H \Psi ^2 \big(28 G^2 H + 14 G H^2 + 7 \dot{G} H + 7 G \dot{H} + 3 H^3 + 5 \dot{H} H + \ddot{H} \big) \notag
  \\
  &\quad\,
  + 6 \alpha \Psi \big(21 G^2 H + 21 G H^2 + 7 G \dot{H} + 3 H^3 + 2 \dot{H} H \big) 
  + 6 H \alpha \dot{\Psi} \big( 7 G + 2 H \big) \label{eq:actionof2ndO}
  \\
  &\quad\,
  + 42 \alpha \bar{\Psi} \big(7 G^3 + 7 G^2 H + G H^2 + \dot{G} H + 2 \dot{G}  G \big) 
  + 42 G \alpha \dot{\bar{\Psi}} \big( 2 G + H \big) \notag
  \\
  &\quad\,
  + \frac{k^2}{a^2} \Big\{ 2 H^2 \Psi^2 + 42 G^2 \bar{\Psi }^2 + 14 G \bar{\Psi } \alpha
  + 4 H \Psi \alpha - 2 \big(7 G + 2 H \big) \chi \alpha + 28 G H \Psi \bar{\Psi } \notag
  \\
  &\quad\, 
  + 14 \big(G^2 - G H + \dot{G} \big) \bar{\Psi } \chi 
  + 14 G \dot{\bar{\Psi }} \chi + 4 \dot{H} \Psi \chi + 4 H \dot{\Psi} \chi \Big\} 
  + \Gamma \mathcal{L}^{(2,1)}_\text{pt} (\Upsilon , H ,G) \Big],
\notag
\end{alignat}
where $S^{(2,0)}_\text{pt}$ represents the leading order part of $\Gamma$ expansion in the second order terms with respect to the scalar perturbations.
Similarly $S^{(2,1)}_\text{pt}$ or $\mathcal{L}^{(2,1)}_\text{pt} (\Upsilon , H ,G)$ does linear order part of $\Gamma$.
The explicit forms of $\mathcal{L}^{(2,1)}_\text{pt} $ and other complicated equations in this section can be found in ref.~\cite{Mathematicacodes}.
By varying the above action, we obtain following equations of motion for scalar perturbations.
\begin{alignat}{3}
  E_{\alpha} &= \frac{k^2}{a^2} \Big\{ 7 G \bar{\Psi } - (7 G + 2 H) \chi + 2 H \Psi \Big\} \notag
  \\
  &\quad\,
  + 3 \Psi \big( 21 G^2 H + 21 G H^2 + 7  G \dot{H} + 3 H^3 + 2 \dot{H} H \big) 
  + 3 H \dot{\Psi} ( 7 G + 2 H ) \label{eq:Ealpha}
  \\
  &\quad\, 
  + 21 \bar{\Psi} \big( 7 G^3 + 7 G^2 H + G H^2 + \dot{G}  H + 2 \dot{G} G \big) 
  + 21 G \dot{\bar{\Psi }} ( 2 G + H ) \notag
  \\
  &\quad\,
  + \Gamma S_\alpha (\Upsilon , H ,G) = 0, \notag
  \\[0.2cm]
  E_{\chi} &= - \alpha  (7 G + 2 H) + 2 \dot{H} \Psi  + 2 H \dot{\Psi }  
  + 7 \bar{\Psi } ( G^2 - G H + \dot{G} ) + 7 G \dot{\bar{\Psi }} \notag
  \\
  &\quad\,
  + \Gamma S_{\chi} (\Upsilon , H ,G) = 0, \label{eq:Echi}
  \\[0.2cm]
  E_{\Psi} &= \frac{k^2}{a^2} \Big\{ 14 G \bar{\Psi }+2 \alpha -2 \chi  (7 G+H)+2 H \Psi -2 \dot{\chi } \Big\}
  - 3 (7 G+2 H) \dot{\alpha } \notag
  \\
  &\quad\,
  + 6 \Psi \big( 28 G^2 H+14 G H^2+7 \dot{G} H+7 G \dot{H}+3 H^3+\ddot{H}+5 \dot{H} H \big) \notag
  \\
  &\quad\, 
  + 21 \bar{\Psi } \big(28 G^3+14 G^2 H+3 G H^2+2 G \dot{H}+2 \dot{G} H+\ddot{G}+15 \dot{G} G \big) \label{eq:EPsi}
  \\
  &\quad\, 
  + 6 \dot{\Psi } (7 G H+3 H^2 + 2 \dot{H} ) + 42 \dot{\bar{\Psi }} (4 G^2 + G H + \dot{G} ) 
  + 6 H \ddot{\Psi } + 21 G \ddot{\bar{\Psi }} \qquad\;\; \notag
  \\
  &\quad\,
  + \Gamma S_{\Psi} (\Upsilon , H ,G) = 0, \notag
\end{alignat}
\begin{alignat}{3}
  E_{\bar{\Psi}} &= \frac{k^2}{a^2} \Big\{ 6 G \bar{\Psi } + \alpha - 2 \chi  (3 G+H)+2 H \Psi -\dot{\chi } \Big\}
  - 3 \dot{\alpha } (2 G+H) \notag
  \\
  &\quad\,
  + 3 \Psi (21 G^2 H+18 G H^2+6 \dot{G} H+6 G \dot{H}+6 H^3+\ddot{H}+7 \dot{H} H ) \notag
  \\
  &\quad\, 
  + 6 \bar{\Psi } (21 G^3+18 G^2 H+6 G H^2+3 G \dot{H}+3 \dot{G} H+\ddot{G}+13 \dot{G} G ) \label{eq:EbarPsi}
  \\
  &\quad\,
  + 6 \dot{\Psi } (3 G H+2 H^2+\dot{H})
  + 6 \dot{\bar{\Psi }} (7 G^2+3 G H+2 \dot{G})
  + 3 H \ddot{\Psi } + 6 G \ddot{\bar{\Psi }} \notag
  \\
  &\quad\,
  + \Gamma S_{\bar{\Psi}} (\Upsilon , H ,G) = 0, \notag
\end{alignat}
where $S_I\,(I=\alpha,\chi,\Psi,\bar{\Psi})$ represents linear terms with respect to $\Gamma$.
At least at the leading order, it is straightforward to show that these are equivalent to the equations of motion (\ref{eq:E12})-(\ref{eq:E72}).
For example, $E_{\Psi}$ is expressed by a combination of $E_2,\dot{E_2}$ and $\frac{k^2}{a^2}E_3$.
Thus the effective action (\ref{eq:actionof2ndO}) is consistent with the results in the previous section.

It is also possible to express the effective action in terms of $P_0$ and $P_1$.
First by eliminating auxiliary fields $\alpha_0$ and $\chi_0$ by using the eq.~(\ref{eq:alpha0}) and the eq.~(\ref{eq:chi0}), 
the leading order action $S^{(2,0)}_\text{pt}$ is written as
\begin{alignat}{3}
  S^{(2,0)}_\text{pt} &= \frac{1}{2 \kappa_{11}^2} \int dt d^3x d^7y \, 6 a_0^3 b_0^7 \, H_0^2 
  \Big\{ - \Big( \frac{k^2}{a_0^2} - \frac{\sqrt{21}-11}{2} H_0^2 \Big) P_0^2 +\dot{P}_0^2 \Big\}.
\end{alignat}
It is easy to confirm that the eq.~(\ref{eq:P0}) can be derived from this action.

Next let us express $S^{(2)}_\text{pt}$ in terms of $P_0$ and $P_1$.
By eliminating auxiliary fields $\alpha$ and $\chi$ by inserting the eq.~(\ref{eq:alpha}) and the eq.~(\ref{eq:chi}) 
into the action (\ref{eq:actionof2ndO}), we obtain
\begin{alignat}{3}
  S^{(2)}_\text{pt} &= \frac{1}{2 \kappa_{11}^2} \int dt d^3x d^7y \, 6 a_0^3 b_0^7  H_0^2  \bigg[
  - \Big( \frac{k^2}{a_0^2} - \frac{\sqrt{21}-11}{2} H_0^2 \Big) P_0^2 + \dot{P_0}{}^2 \notag
  \\
  &\quad\,
  + \Gamma \bigg\{ - 2 \ddot{P_0} P_1 + (\sqrt{21}-1) H_0 \dot{P_0} P_1 + (\sqrt{21}-11) H_0^2 P_0 P_1 \notag
  \\
  &\quad\,
  - \tfrac{3072 (2261 \sqrt{21}-23271) }{1225} H_0^4 \ddot{P_0}^2
  + \big( 3 \bar{a}_1 + 7 \bar{b}_1 + \tfrac{768 (6043049 \sqrt{21}-44300079)}{8575} H_0^6 \big) \dot{P_0}^2 \notag
  \\
  &\quad\, 
  + \big( \tfrac{ \sqrt{21}-11 }{2} (3 \bar{a}_1 + 7 \bar{b}_1 ) 
  + \tfrac{768(16193431 \sqrt{21}-29555691) }{1225} H_0^6 \big) H_0^2 P_0^2 \notag
  \\
  &\quad\,
  + \frac{k^2}{a_0^2} \Big( - 2 P_0 P_1
  - \big( \bar{a}_1 + 7 \bar{b}_1 + \tfrac{ 768 (1807903 \sqrt{21}-20672673) }{8575} H_0^6 \big) P_0^2 \notag
  \\
  &\quad\,
  + \tfrac{1536 (9479 \sqrt{21}-66369) }{1225} H_0^4 \dot{P_0}^2 \Big)
  - \frac{k^4}{a_0^4} \tfrac{6144 (1633 \sqrt{21}-9288) }{1225} H_0^4 P_0^2 \bigg\} \bigg], \label{eqaction2-0}
\end{alignat}
where  $\bar{b}_1$ is defined as
\begin{alignat}{3}
  \bar{b}_1 = - \frac{1+\sqrt{21}}{60} c_g \frac{H_\text{I}^6}{\tau^6}
  = - \frac{1+\sqrt{21}}{60} c_g H_0^6. \label{b1bar}
\end{alignat}
Note that $\dot{\bar{a}}_1$ and $\dot{\bar{b}}_1$ are written as
\begin{alignat}{3}
  \dot{\bar{a}}_1 = H_1 
  =c_h \frac{H_\text{I}^7}{\tau^7}
  =c_h  H_0^7,
  \qquad
\dot{\bar{b}}_1 = G_1 
  =c_g \frac{H_\text{I}^7}{\tau^7}
  =c_g  H_0^7,
\end{alignat}
respectively.
Then it is possible to check that this action consistently reproduces the equations of motion (\ref{eq:P0}) and (\ref{eq:P1}).


\section{Analyses of Scalar Perturbations} \label{sec:analyses}


In this section, first we solve the eq.~(\ref{eq:P0}), and then do the eq.~(\ref{eq:P1}).
From these, it is possible to obtain the explicit form of the curvature perturbation $\psi$
and estimate its power spectrum.


\subsection{Solutions of $P_0$ and $P_1$}


In order to solve the eq.~(\ref{eq:P0}), we introduce new time coordinate $\eta$ instead of $t$, 
which is defined by $dt = \frac{1+\sqrt{21}}{10 H_\text{I}} d\tau = a_0 d\eta$.
Note that $a_0 = a_\text{E} \tau^{\frac{1+\sqrt{21}}{10}}$ is leading part of the scale factor $a$. 
Therefore $\eta$ is considered to be a conformal time after the inflationary expansion.
Then $\eta$ is expressed in terms of $\tau$ like
\begin{alignat}{3}
  \eta &= \frac{1+\sqrt{21}}{10 H_\text{I}} \int \frac{d\tau}{a_0} 
  = \frac{3+\sqrt{21}}{6 a_\text{E} H_\text{I}} \tau^\frac{9-\sqrt{21}}{10}. \label{eq:conftime}
\end{alignat}
By solving inversely, $\tau$ is expressed in terms of $\eta$ as
\begin{alignat}{3}
  \tau &= \Big( \frac{\sqrt{21}-3}{2} a_\text{E} H_\text{I} \eta \Big)^\frac{9+\sqrt{21}}{6}, \label{eq:taueta}
\end{alignat}
and $a_0$ is given by
\begin{alignat}{3}
  a_0 = a_\text{E} \Big( \frac{\sqrt{21}-3}{2} a_\text{E} H_\text{I} \eta \Big)^\frac{3+\sqrt{21}}{6}. \label{eq:a0conftime}
\end{alignat}
Remind that the inflationary expansion is realized during $1 \leq \tau \leq 2$.
Now $\tau = 1$ corresponds to $a_\text{E} H_\text{I} \eta = \frac{3+\sqrt{21}}{6} \sim 1.26$, and
$\tau = 2$ does to $a_\text{E} H_\text{I} \eta = \frac{3+\sqrt{21}}{6} 2^\frac{9-\sqrt{21}}{10} \sim 1.72$.

Hubble parameter $\mathcal{H}_0$ with respect to the time coordinate $\eta$ is evaluated as
\begin{alignat}{3}
  \mathcal{H}_0 = \frac{a_0'}{a_0} = a_0 H_0 = \frac{3+\sqrt{21}}{6} \frac{1}{\eta}, \label{eq:mcH0}
\end{alignat}
where $'=\frac{d}{d\eta}$.
And derivatives of $P_0$ with respect to $t$ are replaced with
\begin{alignat}{3}
  \dot{P}_0 &= \frac{1}{a_0} P'_0, \qquad \ddot{P}_0 = \frac{1}{a_0^2} ( P''_0 - \mathcal{H}_0 P'_0). \label{eq:difP0-1}
\end{alignat}
Then by multiplying $a_0^2$ to the eq.~(\ref{eq:P0}), it becomes
\begin{alignat}{3}
  0 &= P_0'' - \frac{\sqrt{21}+1}{2} \mathcal{H}_0 P_0' + \Big( k^2 - \frac{\sqrt{21}-11}{2} \mathcal{H}_0^2 \Big) P_0 \notag
  \\
  &= a_0^{\frac{\sqrt{21}+1}{4}} \bigg[ U_0'' + \Big(k^2 + \frac{1}{4 \eta^2} \Big) U_0 \bigg], \label{eq:U0}
\end{alignat}
where $P_0$ and $U_0$ are related as
\begin{alignat}{3}
  P_0 = a_0^{\frac{\sqrt{21}+1}{4}} U_0. \label{eq:defU0}
\end{alignat}
Now the eq.~(\ref{eq:U0}) can be solved as
\begin{alignat}{3}
  U_0 = c_1 \sqrt{k \eta} J_0(k \eta) + c_2 \sqrt{k \eta} Y_0(k \eta). \label{eq:solU0}
\end{alignat}
$J_0$ and $Y_0$ are Bessel functions of the first and second kind, respectively.
$c_1$ and $c_2$ are integral constants and both have mass dimension $-1$.
In order to fix the ratio of $c_2/c_1$, we demand that $U_0$ behaves like $e^{-ik\eta}$ as $\eta$ goes to the infinity.
This is reasonable if we assume that the perturbations, such as the eq.~(\ref{eq:FmPsi}), 
are canonically expressed in terms of Fourier modes as $\eta$ goes to the infinity.
Since $\sqrt{x} J_0(x) \sim \sqrt{\frac{2}{\pi}} \cos(x-\frac{\pi}{4})$ and $\sqrt{x} Y_0(x) \sim \sqrt{\frac{2}{\pi}} \sin(x-\frac{\pi}{4})$
as $x \to \infty$, we choose $\frac{c_2}{c_1}$ as
\begin{alignat}{3}
  \frac{c_2}{c_1} = -i, \label{eq:c2/c1}
\end{alignat}
and $U_0$ is given by
\begin{alignat}{3}
  U_0 = c_1 \sqrt{k \eta} H^{(2)}_0(k \eta), \label{eq:solU02}
\end{alignat}
where $H^{(2)}_0$ is Hankel function of the second kind.

Next let us solve the differential equation for $P_1$.
We replace derivatives of $P_0$ with respect to $t$ by using the eq.~(\ref{eq:difP0-1}) and 
\begin{alignat}{3}
  \dddot{P}_0 &= \frac{1}{a_0^3} \Big\{ P'''_0 - 3 \mathcal{H}_0 P''_0 + \frac{1+\sqrt{21}}{2} \mathcal{H}_0^2 P'_0 \Big\}, \label{eq:difP0-2}
  \\
  \ddddot{P}_0 &= \frac{1}{a_0^4} \Big\{ P''''_0 - 6 \mathcal{H}_0 P'''_0 
  + (5+2\sqrt{21}) \mathcal{H}_0^2 P''_0 - \frac{21+\sqrt{21}}{2} \mathcal{H}_0^3 P'_0 \Big\}. \notag
\end{alignat}
Then by multiplying $a_0^2$ to the eq.~(\ref{eq:P1}), it becomes
\begin{alignat}{3}
  &P_1'' - \frac{\sqrt{21}+1}{2} \mathcal{H}_0 P_1' 
  + \Big( k^2 - \frac{\sqrt{21}-11}{2} \mathcal{H}_0^2 \Big) P_1 \notag
  \\
  &\! + \frac{1}{a_0^6} \Big[ -\tfrac{1536 (49692383 \sqrt{21}-70593438) }{8575} \mathcal{H}_0^8 P_0 
  + \tfrac{1536 (15053494 \sqrt{21}-40585737) }{1715} \mathcal{H}_0^7 P_0' \notag
  \\
  &\! + \tfrac{6912 (1812421 \sqrt{21}-16802761) }{8575} \mathcal{H}_0^6 P_0''
  + \tfrac{6144 (57047 \sqrt{21}-107067) }{1225} \mathcal{H}_0^5 P'''_0
  + \tfrac{3072 (2261 \sqrt{21}-23271) }{1225} \mathcal{H}_0^4 P''''_0 \notag
  \\
  &\! + k^2 \Big\{ \big( \tfrac{768 (3567079 \sqrt{21}-29260239) }{8575} \mathcal{H}_0^6 - 2 a_0^6 \bar{a}_1 \big) P_0
  + \tfrac{1536 (161183 \sqrt{21}-464463) }{1225} \mathcal{H}_0^5 P_0' \notag
  \\
  &\! + \tfrac{1536 (9479 \sqrt{21}-66369) }{1225} \mathcal{H}_0^4 P_0'' \Big\} 
  + k^4 \tfrac{6144 (1633 \sqrt{21}-9288) }{1225} \mathcal{H}_0^4 P_0 \Big] = 0. \label{eq:P1-2}
\end{alignat}
In the above, we used
\begin{alignat}{3}
  \mathcal{H}'_0 = \frac{3-\sqrt{21}}{2} \mathcal{H}_0^2, \qquad 
  \mathcal{H}''_0 = 3(5-\sqrt{21}) \mathcal{H}_0^3, \label{eq:difmcH}
\end{alignat}
and neglected higher order terms on $\Gamma$.
The remaining eq.~(\ref{eq:E72}) is automatically satisfied. 
If we rescale $P_1$ as
\begin{alignat}{3}
  P_1 = a_0^{\frac{\sqrt{21}+1}{4}} U_1, \label{eq:defU1}
\end{alignat}
we obtain
\begin{alignat}{3}
  0 &= U_1'' + \Big( k^2 -\frac{3(\sqrt{21}-5)}{8} \mathcal{H}_0^2 \Big) U_1 \notag
  \\
  &\quad
  + \frac{1}{a_0^6} \Big[ - \tfrac{3456 (64904378 \sqrt{21}-301020693) }{8575} \mathcal{H}_0^8 U_0
  + \tfrac{13824 (676051 \sqrt{21}-2068971) }{1715} \mathcal{H}_0^7 U_0' \notag
  \\
  &\quad
  + \tfrac{2304 (4137503\sqrt{21}-32998023) }{8575} \mathcal{H}_0^6 U_0''
  + \tfrac{36864 (7757\sqrt{21}-15827) }{1225} \mathcal{H}_0^5 U_0'''
  + \tfrac{1536 (4522 \sqrt{21}-46542) }{1225} \mathcal{H}_0^4 U_0'''' \notag
  \\
  &\quad
  + k^2 \Big\{ \big( \tfrac{2304 (680348 \sqrt{21}-5726943) }{8575} \mathcal{H}_0^6 - 2 a_0^6 \bar{a}_1 \big) U_0
  + \tfrac{9216 (22123 \sqrt{21}-66353) }{1225} \mathcal{H}_0^5 U_0' \notag
  \\
  &\quad
  + \tfrac{1536 (9479 \sqrt{21}-66369) }{1225} \mathcal{H}_0^4 U_0'' \Big\}
  + k^4 \tfrac{6144 (1633 \sqrt{21}-9288) }{1225} \mathcal{H}_0^4 U_0 \Big] \notag
  \\[0.1cm]
  &= U_1'' + \Big( k^2 + \frac{1}{4 \eta ^2} \Big) U_1
  + \frac{ 2^8}{3 \cdot 5^2 \, 7^3} \Big( \frac{\sqrt{21}+3}{6} \Big)^{3+\sqrt{21}}
  \frac{\sqrt{k \eta}}{a_\text{E}^6 \eta^8 (a_\text{E} H_\text{I} \eta)^{3+\sqrt{21}} } \notag
  \\
  &\quad\,
  \Big[ \big( 420 (127267+27753 \sqrt{21}) k^3 \eta^3
  - 78 (27149229+5923661 \sqrt{21}) k \eta \big) \big( c_1 J_1(k \eta) + c_2 Y_1(k \eta) \big) \notag
  \\
  &\quad\,
  + \big( -44100 (37+8\sqrt{21}) k^4 \eta^4 + (602616417+131422261 \sqrt{21}) k^2 \eta^2 \label{eq:U1}
  \\
  &\quad\,
  + 36 (49428849+10780291 \sqrt{21}) \big) \big( c_1 J_0(k \eta) + c_2 Y_0(k \eta) \big) \Big]. \notag
\end{alignat}
In the last line, we substituted the eq.~(\ref{eq:mcH0}) and the eq.~(\ref{eq:solU0}).
Particular solution of the above is given by
\begin{alignat}{3}
  U_1 &= - \tfrac{288 (20727 - 4523 \sqrt{21})}{300125} 
  \frac{H_\text{I}^6 \sqrt{k \eta}}{\big( \frac{\sqrt{21}-3}{2} a_\text{E} H_\text{I} \eta \big)^{9+\sqrt{21}}} 
  \Big( c_1 U_{11} 
  - c_2 \tfrac{(- 41 + 9 \sqrt{21}) \sqrt{\pi}}{10} U_{12} \Big). \label{eq:U1sol}
\end{alignat}
Since the explicit expressions of $U_{11}$ and $U_{12}$ are quite long, we put them in the appendix \ref{sec:supp}.
The ratio of $\frac{c_2}{c_1}$ should be fixed by $\frac{c_2}{c_1} = -i$ as explained around the eq.~(\ref{eq:c2/c1}).
Thus we have solved the eqs.(\ref{eq:P0}) and (\ref{eq:P1}), and $P = P_0 + \Gamma P_1$ is given by
\begin{alignat}{3}
  P_0 &= c_1 a_0^{\frac{\sqrt{21}+1}{4}} \sqrt{k \eta} H^{(2)}_0(k \eta), \label{eq:solP0,P1}
  \\
  P_1 &= - \tfrac{288 (20727 - 4523 \sqrt{21})}{300125} c_1 a_0^{\frac{\sqrt{21}+1}{4}}
  \frac{H_\text{I}^6 \sqrt{k \eta}}{\big( \frac{\sqrt{21}-3}{2} a_\text{E} H_\text{I} \eta \big)^{9+\sqrt{21}}} 
  \Big( U_{11} + i \tfrac{(- 41 + 9 \sqrt{21}) \sqrt{\pi}}{10} U_{12} \Big). \notag
\end{alignat}


\subsection{Curvature perturbation}


Let us investigate the curvature perturbation $\psi$.
If we choose $\bar{\psi} = 0$ gauge, the curvature perturbation is expressed as $\psi = HP$.
Up to the linear order of $\Gamma$ expansion, $\psi$ is written as
\begin{alignat}{3}
  \psi(\eta,x) = \psi_0 + \Gamma \psi_1 = H_0 P_0 + \Gamma (H_1 P_0 + H_0 P_1), \label{eq:psidef}
\end{alignat}
where $\eta$ is related to $\tau$ via the eq.~(\ref{eq:conftime}).
Now we expand $\psi(\eta,x)$ as
\begin{alignat}{3}
  \psi(\eta,x) &= \int d^3k \, \big\{ \psi(\eta,k) e^{i k_i x^i} + \psi(\eta,k)^\ast e^{-i k_i x^i} \big\}. \label{eq:Fmpsi}
\end{alignat}
Then Fourier component of $\psi_0$ is evaluated as
\begin{alignat}{3}
  \psi_0(\eta,k) &= \tilde{c}_1 H^{(2)}_0(k \eta), \qquad
  \tilde{c}_1 &= c_1 a_\text{E}^{\frac{\sqrt{21}+1}{4}} H_\text{I} 
  \Big( \frac{\sqrt{21}+3}{6} \frac{k}{a_\text{E} H_\text{I}} \Big)^\frac{1}{2}. \label{eq:psi0}
\end{alignat}
If we take $k\eta \to \infty$, $\psi_0(\eta,k)$ approaches to $\tilde{c}_1 \sqrt{\frac{2}{\pi k\eta}} e^{i\frac{\pi}{4}} e^{-ik\eta}$.

Next Fourier component of $\psi_1$ is calculated as
\begin{alignat}{3}
  \psi_1(\eta,k) &= H_0 (c_h H_0^6 P_0 + P_1) \notag
  \\
  &= c_1 a_0^{\frac{\sqrt{21}+1}{4}} H_0^7 \sqrt{k \eta} \Big\{ c_h H^{(2)}_0(k \eta) 
  - \tfrac{288 (20727 - 4523 \sqrt{21})}{300125} \big( U_{11} + i \tfrac{(- 41 + 9 \sqrt{21}) \sqrt{\pi}}{10} U_{12} \big) \Big\} \notag
  \\
  &= \frac{ \tilde{c}_1 H_\text{I}^6 }{ \tau^6 } \Big\{ c_h H^{(2)}_0(k \eta) 
  - \tfrac{288 (20727 - 4523 \sqrt{21})}{300125} \big( U_{11} + i \tfrac{(- 41 + 9 \sqrt{21}) \sqrt{\pi}}{10} U_{12} \big) \Big\}. \label{eq:psi1}
\end{alignat}
Here we used $\tau^6 = ( \frac{\sqrt{21}-3}{2} a_\text{E} H_\text{I} \eta )^{9+\sqrt{21}}$. 
Note that $\psi_1$ decreases faster than $\psi_0$ as $\tau$ goes to the infinity.

Finally the power spectrum of the Fourier mode up to linear order of $\Gamma$ is expressed as
\begin{alignat}{3}
  \mathcal{P}(\eta,k) &= \Big( \frac{k}{a_\text{E} H_\text{I}} \Big)^3 | \psi(\eta,k) |^2 \notag
  \\
  &= |\tilde{c}_1|^2 \Big( \frac{k}{a_\text{E} H_\text{I}} \Big)^3 \Big| H_0^{(2)}(\eta,k) 
  + \frac{ \Gamma H_\text{I}^6 }{ \tau^6 } \Big\{ c_h H_0^{(2)}(\eta,k) \notag
  \\
  &\quad\,
  - \tfrac{288 (20727 - 4523 \sqrt{21})}{300125} 
  \big( U_{11}(k \eta) + i \tfrac{(- 41 + 9 \sqrt{21}) \sqrt{\pi}}{10} U_{12}(k \eta) \big) \Big\} \Big|^2. \label{eq:PS}
\end{alignat}
Now we assume that the power spectrum was some constant $A$ at the beginning of the universe, $\tau = 1$ 
or $\eta = \tfrac{3+\sqrt{21}}{6 a_\text{E} H_\text{I}}$.
Then $\mathcal{P}(\tfrac{3+\sqrt{21}}{6 a_\text{E} H_\text{I}},k) = A$, and
$k$ dependence of $|\tilde{c_1}|$ is given by
\begin{alignat}{3}
  |\tilde{c}_1|^2 &= A \Big( \frac{a_\text{E} H_\text{I}}{k} \Big)^3
  \Big| H_0^{(2)}(\tfrac{3+\sqrt{21}}{6 a_\text{E} H_\text{I}}, k) 
  + \frac{ \Gamma H_\text{I}^6 }{ \tau^6 } \Big\{ c_h H_0^{(2)}(\tfrac{3+\sqrt{21}}{6 a_\text{E} H_\text{I}}, k) \notag
  \\
  &\quad\,
  - \tfrac{288 (20727 - 4523 \sqrt{21})}{300125} 
  \big( U_{11} (\tfrac{3+\sqrt{21}}{6 a_\text{E} H_\text{I}} k) 
  + i \tfrac{(- 41 + 9 \sqrt{21}) \sqrt{\pi}}{10} U_{12} (\tfrac{3+\sqrt{21}}{6 a_\text{E} H_\text{I}}k) \big) 
  \Big\} \Big|^{-2}. \label{eq:c1tilde}
\end{alignat}

\begin{figure}[htb]
 \centering
 \begin{picture}(240,185)
 \put(273,153){$a_\text{E} H_\text{I} \eta$}
 \put(-35,0){$\log \frac{\mathcal{P}}{A}$}
 \includegraphics[keepaspectratio,scale=0.75]{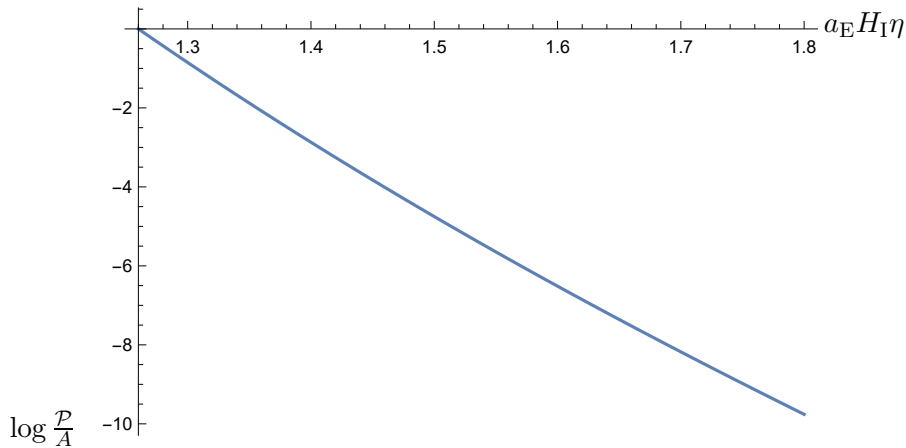}
 \end{picture}
 \caption{Plot of the power spectrum with $\log \frac{k}{a_\text{E} H_\text{I}}=-30$ and $\Gamma H_\text{I}^6 = 0.014$.}
 \label{fig:PS1}
\end{figure}

\begin{figure}[htb]
 \centering
 \begin{picture}(240,185)
 \put(-45,165){$\log \frac{k}{a_\text{E} H_\text{I}}$}
 \put(273,0){$\log \frac{\mathcal{P}}{A}$}
 \includegraphics[keepaspectratio,scale=0.75]{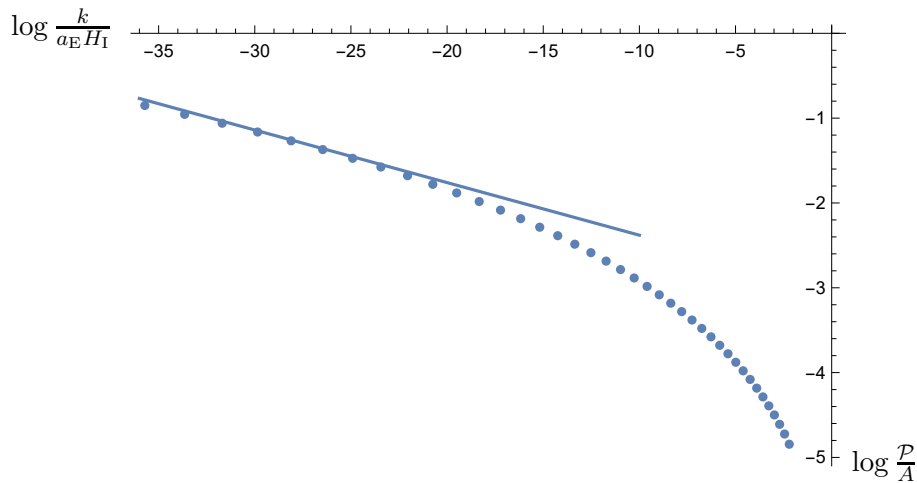}
 \end{picture}
 \caption{Plot of the power spectrum at $k=aH$ with $\Gamma H_\text{I}^6 = 0.014$. The line is given by 
 $\log \frac{\mathcal{P}}{A} = - 0.062 \log \frac{k}{a_\text{E} H_\text{I}} -3.0$.}
 \label{fig:PS2}
\end{figure}

So far the analyses in this paper is reliable up to the linear order of $\Gamma$. 
Basically we don't have any control over higher order terms, because we have poor knowledge about coefficients of those terms.
At best we know that coefficient of those terms behave like $(\Gamma H_\text{I}^6)^n$, 
and expect that the $n=1$ term in this paper is dominant during the inflationary era.
Then the behavior of the power spectrum (\ref{eq:PS}) with $\log \frac{k}{a_\text{E} H_\text{I}}=-30$ and $\Gamma H_\text{I}^6 = 0.014$ 
is plotted as in fig.~\ref{fig:PS1}.
The shape of the plot does not change so much even if the wave number ranges $-40 < \log \frac{k}{a_\text{E} H_\text{I}} < -10$.
The power spectrum is monotonically decreasing, but its tilt becomes slightly mild after the inflationary era.
The behavior of the power spectrum at the horizon crossing $k=aH$ with $\Gamma H_\text{I}^6 = 0.014$ is shown in fig.~\ref{fig:PS2}.
If we fit the data between $-36 < \log \frac{k}{a_\text{E} H_\text{I}} < -20$ in fig.~\ref{fig:PS2}, 
it is possible to draw a line $\log \frac{\mathcal{P}}{A} = - 0.062 \log \frac{k}{a_\text{E} H_\text{I}} -3.0$, and
the spectral index is estimated as $n_s = 0.94$. 
This is close to the value of current observation, and we should investigate more seriously the era after the inflation.

\newpage

\section{Conclusion and Discussion}


In this paper we investigated the inflationary scenario in M-theory.
In addition to the ordinary supergravity part, the effective action of the M-theory contains higher curvature terms, 
which are expressed by products of 4 Weyl tensors. 
In the early universe, $H$ in the eq.~(\ref{eq:HGsol}) is relatively large and nonzero components of the Weyl tensor also become large.
So the higher curvature terms become important and those induce the inflationary expansion.
After sufficient expansion, $H$ becomes small and the Weyl tensor does small.
Then the higher curvature terms are negligible and inflation naturally ends.

The main purpose of this paper is to explore the scalar perturbations in the above inflationary scenario.
Actually, we considered the metric perturbations around the homogeneous and isotropic background,
and derived the linearized equations of motion for the scalar perturbations.
Originally there are 4 equations which linearly depend on $\alpha$, $\chi$, $\Psi$ and $\bar{\Psi}$, 
but after eliminating auxiliary fields $\alpha$ and $\chi$, we obtain only one equation for $P=\Psi-\bar{\Psi}$.
The equation is expanded with respect to the parameter $\Gamma$ up to the linear order, 
and the results are given by the eqs.~(\ref{eq:P0}) and (\ref{eq:P1}).

We also constructed the effective action for the scalar perturbations, and
confirmed that the eqs.~(\ref{eq:P0}) and (\ref{eq:P1}) can be reproduced from the effective action for $P$.
Then we solved $P$ up to the linear order of $\Gamma$, as shown in the eq.~(\ref{eq:solP0,P1}), 
and obtained the power spectrum of the curvature perturbation $\psi$.
As an initial condition, we assumed that the power spectrum of the curvature perturbation was some constant.
Then the power spectrum is plotted against the time evolution in the fig.~\ref{fig:PS1}.
The power spectrum is monotonically decreasing, but its tilt becomes mild after the inflationary era.
We also plotted the power spectrum against the wave number in the fig.~\ref{fig:PS2}.
The figure shows that the scalar spectral index becomes $n_s = 0.94$, if we fit the data for the wide range of the wave numbers.
This is close to the value of current observation, and we should investigate more seriously the era after the inflation.

In the above analyses, we neglected the wave number $l_m$ of the internal space.
So the next task is to include it and investigate the effect to the power spectrum.
Of course, the tensor to scalar ratio should be evaluated in the above inflationary scenario.
As a future work, it is interesting to apply the method developed here 
to more complicated internal geometry, such as $G_2$ manifold \cite{Brandhuber:2001yi}. 
It is also interesting to apply the analyses of this paper 
to the heterotic superstring theory with nontrivial internal space, which contains $R^2$ corrections\cite{Brandle:2000qp},
and reveal several problems in string cosmology\cite{Antoniadis:2016avv}. 
Unification of the inflationary expansion and late time acceleration in modified gravity, such as $f(R)$ gravity or mimetic gravity, is an interesting direction 
to be explored\cite{Nojiri:2003rz,Carroll:2003wy,Myrzakulov:2015qaa}.

\section*{Acknowledgement}
The authors would like to thank members in particle physics groups at Ibaraki University, Waseda University and KEK theory center for useful discussions.
We would also like to thank the Yukawa Institute for Theoretical Physics
at Kyoto University for hospitality during the workshop YITP-W-19-08 ``Strings and Fields 2019", where part of this work was carried out.
This work was partially supported by Japan Society for the Promotion of Science, Grant-in-Aid for Scientific Research (C) Grant Number JP17K05405.

\appendix

\section{Supplementary Notes} \label{sec:supp}


The explicit form of $U_{11}$ in the eq.~(\ref{eq:U1sol}) is given by
\begin{alignat}{3}
  &U_{11}(k \eta) &\notag
  \\
  &= \sqrt{\pi} J_0(k \eta) \Big\{ 
  7350 (13881+3029 \sqrt{21}) k^4 \eta^4 G_{3,5}^{2,2} \Big( k\eta ,\tfrac{1}{2} \Big|
  \begin{array}{c}
    \frac{1}{2}, \frac{7+\sqrt{21}}{2}, -\frac{1}{2} \\
    0, 0 , -\frac{1}{2}, 0, \frac{5+\sqrt{21}}{2} \\
  \end{array}
  \Big) \notag
  \\
  &\quad\,
  -1960 (1712457+373688 \sqrt{21}) k^3 \eta^3 G_{3,5}^{2,2} \Big( k\eta, \tfrac{1}{2} \Big|
  \begin{array}{c}
    \frac{1}{2}, \frac{8+\sqrt{21}}{2} , 0 \\
    \frac{1}{2}, \frac{1}{2}, -\frac{1}{2}, -\frac{1}{2}, \frac{6+\sqrt{21}}{2} \\
  \end{array}
  \Big) \notag
  \\
  &\quad\,
  - (37841511589+8257680071 \sqrt{21}) k^2 \eta^2 G_{3,5}^{2,2} \Big( k\eta , \tfrac{1}{2} \Big|
  \begin{array}{c}
    \frac{1}{2}, \frac{9+\sqrt{21}}{2}, -\frac{1}{2} \\
    0, 0, -\frac{1}{2}, 0, \frac{7+\sqrt{21}}{2} \\
  \end{array}
  \Big) \notag
  \\
  &\quad\,
  + 78 (1705246067+372115053 \sqrt{21}) k \eta G_{3,5}^{2,2} \Big( k\eta, \tfrac{1}{2} \Big|
  \begin{array}{c}
    \frac{1}{2}, 5+\frac{\sqrt{21}}{2}, 0 \\
    \frac{1}{2}, \frac{1}{2}, -\frac{1}{2}, -\frac{1}{2}, \frac{8+\sqrt{21}}{2} \\
  \end{array}
  \Big) \notag
  \\
  &\quad\,
  - 72 (1551990251+338671484 \sqrt{21}) G_{3,5}^{2,2} \Big( k\eta, \tfrac{1}{2} \Big|
  \begin{array}{c}
    \frac{1}{2}, \frac{11+\sqrt{21}}{2}, -\frac{1}{2} \\
    0, 0, -\frac{1}{2}, 0, \frac{9+\sqrt{21}}{2} \\
  \end{array} \Big) \Big\} \label{eq:U11}
  \\
  &\quad\,
  + 2\pi Y_0(k \eta) \Big\{
  7350 (1449+316 \sqrt{21}) k^4 \eta^4 \,
  _2F_3 \big(\tfrac{1}{2}, -\tfrac{5+\sqrt{21}}{2}; 1, 1, -\tfrac{3+\sqrt{21}}{2}; - k^2 \eta^2 \big) \notag
  \\
  &\quad\,
  - 245 (714837+155983 \sqrt{21}) k^4 \eta^4 \,
  _2F_3 \big(\tfrac{3}{2}, -\tfrac{5+\sqrt{21}}{2}; 2, 2, -\tfrac{3+\sqrt{21}}{2}; - k^2 \eta^2 \big) \notag
  \\
  &\quad\,
  - (3267117844+712937461 \sqrt{21}) k^2 \eta^2 \,
  _2F_3 \big( \tfrac{1}{2}, -\tfrac{7+\sqrt{21}}{2}; 1, 1, -\tfrac{5+\sqrt{21}}{2}; - k^2 \eta^2 \big) \notag
  \\
  &\quad\,
  + 39 (147225227+32127118 \sqrt{21}) k^2 \eta^2 \,
  _2F_3 \big(\tfrac{3}{2}, -\tfrac{7+\sqrt{21}}{2}; 2, 2, -\tfrac{5+\sqrt{21}}{2}; - k^2 \eta^2 \big) \notag
  \\
  &\quad\,
  - 6 (1371162219+299210621 \sqrt{21}) \,
  _2F_3 \big( \tfrac{1}{2}, -\tfrac{9+\sqrt{21}}{2}; 1, 1, -\tfrac{7+\sqrt{21}}{2}; - k^2 \eta^2 \big) \Big\}. \notag
\end{alignat}
Here the function 
$G_{p,q}^{m,n} \big(z,r|\begin{array}{c} a_1, \cdots, a_n, a_{n+1}, \cdots, a_p \\ b_1, \cdots, b_m, b_{m+1}, \cdots, b_q \\ \end{array} \big)$
is the generalized Meijer G-function, and the function
$_pF_q \big(a_1,\cdots,a_p; b_1,\cdots,b_q; z \big)$ 
is the generalized hypergeometric function.
The explicit form of $U_{12}$ in the eq.~(\ref{eq:U1sol}) is given by
\begin{alignat}{3}
  &U_{12}(k \eta) \notag
  \\
  &= J_0(k \eta) \Big\{ 
  7350 (119133+25997 \sqrt{21}) \sqrt{\pi} k^4 \eta^4 \,
  _2F_3 \big( \tfrac{1}{2},-\tfrac{5+\sqrt{21}}{2};1,1,-\tfrac{3+\sqrt{21}}{2};-k^2 \eta^2 \big) \notag
  \\
  &\quad\,
  - 980 (14697276+3207209 \sqrt{21} ) \sqrt{\pi} k^4 \eta^4 \,
  _2F_3 \big( \tfrac{3}{2},-\tfrac{5+\sqrt{21}}{2};2,2,-\tfrac{3+\sqrt{21}}{2};-k^2 \eta^2 \big) \notag
  \\
  &\quad\,
  - (268697011733+58634496497 \sqrt{21}) \sqrt{\pi} k^2 \eta^2 \,
  _2F_3 \big(\tfrac{1}{2},-\tfrac{7+\sqrt{21}}{2};1,1,-\tfrac{5+\sqrt{21}}{2};-k^2 \eta^2 \big) \notag
  \\
  &\quad\,
  + 39 (12108259609+2642238881 \sqrt{21}) \sqrt{\pi} k^2 \eta^2 \,
  _2F_3 \big( \tfrac{3}{2},-\tfrac{7+\sqrt{21}}{2};2,2,-\tfrac{5+\sqrt{21}}{2};-k^2 \eta^2 \big) \notag
  \\
  &\quad\,
  - 24 (28192114587+6152023858 \sqrt{21}) \sqrt{\pi} \,
  _2F_3 \big( \tfrac{1}{2},-\tfrac{9+\sqrt{21}}{2};1,1,-\tfrac{7+\sqrt{21}}{2};-k^2 \eta^2 \big) \notag
  \\
  &\quad\,
  -14700 (570801+124559 \sqrt{21}) k^4 \eta^4 G_{3,5}^{3,1} \Big( k\eta, \tfrac{1}{2} \Big|
  \begin{array}{c}
    \tfrac{7+\sqrt{21}}{2}, \tfrac{1}{2}, \tfrac{1}{2} \\
    0, 0, 0, \tfrac{1}{2}, \tfrac{5+\sqrt{21}}{2} \\
  \end{array} \Big) \label{eq:U12}
  \\
  &\quad\,
  + 1960 (140837769+30733321 \sqrt{21}) k^3 \eta^3 G_{3,5}^{3,1} \Big( k\eta, \tfrac{1}{2} \Big|
  \begin{array}{c}
    \tfrac{8+\sqrt{21}}{2} , 0, 0 \\
    -\tfrac{1}{2}, -\tfrac{1}{2}, \tfrac{1}{2}, 0, \tfrac{6+\sqrt{21}}{2} \\
  \end{array} \Big) \notag
  \\
  &\quad\,
  + 4 (778050877142+169784621803 \sqrt{21}) k^2 \eta^2 G_{3,5}^{3,1} \Big( k\eta, \tfrac{1}{2} \Big|
  \begin{array}{c}
    \tfrac{9+\sqrt{21}}{2}, \tfrac{1}{2}, \tfrac{1}{2} \\
    0, 0, 0, \tfrac{1}{2}, \tfrac{7+\sqrt{21}}{2} \\
  \end{array} \Big) \notag
\end{alignat}
\begin{alignat}{3}
  &\quad\,
  - 312 (35061208441+7650982944 \sqrt{21}) k \eta G_{3,5}^{3,1} \Big( k\eta, \tfrac{1}{2} \Big|
  \begin{array}{c}
    \tfrac{10+\sqrt{21}}{2}, 0, 0 \\
    - \tfrac{1}{2}, - \tfrac{1}{2}, \tfrac{1}{2}, 0, \tfrac{8+\sqrt{21}}{2} \\
  \end{array} \Big) \qquad\qquad \notag
  \\
  &\quad\,
  + 72 (127640510767+27853443103 \sqrt{21}) G_{3,5}^{3,1} \Big( k\eta, \tfrac{1}{2} \Big|
  \begin{array}{c}
    \tfrac{11+\sqrt{21}}{2} , \tfrac{1}{2}, \tfrac{1}{2} \\
    0, 0, 0, \tfrac{1}{2}, \tfrac{9+\sqrt{21}}{2} \\
  \end{array} \Big) \Big\} \notag
  \\
  &\quad\,
  + Y_0(k\eta) \Big\{ 
  7350 (570801+124559 \sqrt{21}) k^4 \eta^4 G_{3,5}^{2,2} \Big( k\eta , \tfrac{1}{2} \Big|
  \begin{array}{c}
    \tfrac{1}{2}, \tfrac{7+\sqrt{21}}{2}, - \tfrac{1}{2} \\
    0, 0, - \tfrac{1}{2}, 0, \tfrac{5+\sqrt{21}}{2} \\
  \end{array} \Big) \notag
  \\
  &\quad\,
  - 980 (140837769+30733321 \sqrt{21}) k^3 \eta^3 G_{3,5}^{2,2} \Big( k\eta , \tfrac{1}{2} \Big|
  \begin{array}{c}
    0, \tfrac{8+\sqrt{21}}{2}, -1 \\
    - \tfrac{1}{2}, \tfrac{1}{2}, -1, -\tfrac{1}{2}, \tfrac{6+\sqrt{21}}{2} \\
  \end{array} \Big) \notag
  \\
  &\quad\,
  -2 (778050877142+169784621803 \sqrt{21}) k^2 \eta^2 G_{3,5}^{2,2} \Big( k\eta, \tfrac{1}{2} \Big|
  \begin{array}{c}
    \tfrac{1}{2}, \tfrac{9+\sqrt{21}}{2}, -\tfrac{1}{2} \\
    0, 0, -\tfrac{1}{2}, 0, \tfrac{7+\sqrt{21}}{2} \\
  \end{array} \Big) \notag
  \\
  &\quad\,
  + 156 (35061208441+7650982944 \sqrt{21}) k \eta G_{3,5}^{2,2} \Big( k\eta, \tfrac{1}{2} \Big|
  \begin{array}{c}
    0, \tfrac{10+\sqrt{21}}{2}, -1 \\
    - \tfrac{1}{2}, \tfrac{1}{2}, -1, -\tfrac{1}{2}, \tfrac{8+\sqrt{21}}{2} \\
  \end{array} \Big) \notag
  \\
  &\quad\,
  - 36 (127640510767+27853443103 \sqrt{21}) G_{3,5}^{2,2} \Big( k\eta, \tfrac{1}{2} \Big|
  \begin{array}{c}
    \tfrac{1}{2}, \tfrac{11+\sqrt{21}}{2}, -\tfrac{1}{2} \\
    0, 0, - \tfrac{1}{2}, 0, \tfrac{9+\sqrt{21}}{2} \\
  \end{array} \Big) \Big\}. \notag
\end{alignat}


\section{Dimensional Reduction} \label{sec:appB}


In this appendix, we consider dimensional reduction to 4 dimensional spacetime in Einstein frame.
Let us denote $g_{\mu\nu}^\text{E}$ as the metric in 4 dimensional Einstein frame.
The 11 dimensional metric is written in terms of $g_{\mu\nu}^\text{E}$ as
\begin{alignat}{3}
  ds^2 = e^{-7\rho} (g^\text{E}_{\mu\nu} dx^\mu dx^\nu) + e^{2\rho} dy_m^2, \label{eq:met4d}
\end{alignat}
where $e^\rho = b$ and $\mu,\nu=0,1,2,3$.
Then 4 dimensional metric is written by
\begin{alignat}{3}
  ds_4^2 = g^\text{E}_{\mu\nu} dx^\mu dx^\nu = - b^7 dt^2 + a^2 b^7 dx_i^2 = a^2 b^7 \Big( - \Big( \frac{dt}{a} \Big)^2 + dx_i^2 \Big).
\end{alignat}
The scale factor in 4 dimensional spacetime is defined as $\tilde{a} = a b^{7/2}$, but
the particle horizon is the same as the eqs.~(\ref{eq:ph1}) and (\ref{eq:ph2}).
So if the inflation occurs in 11 dimensions, it is also true in 4 dimensions.
Notice also that the curvature perturbation $\psi$ in the eq.~(\ref{eq:scalarpt}) is the same as that in 4 dimensions
since it is defined as a multiplicative factor of the background scale factor.
Thus the analysis on the scalar spectrum index still holds in 4 dimensions.

Let us examine the behavior of the scale factor $\tilde{a}$ in 4 dimensions.
From the eq.~(\ref{eq:abQc}), it is explicitly written as
\begin{alignat}{3}
  \log \frac{\tilde{a}}{\tilde{a}_\text{E}} &= \frac{9-\sqrt{21}}{20} \log \tau 
  + \frac{1+\sqrt{21}}{60} \frac{(- c_h - \frac{7}{2} c_g) \Gamma H_\text{I}^6}{\tau^6}, \qquad
  - c_h - \tfrac{7}{2} c_g \sim 15873.
\end{alignat}
$a_\text{E}$ corresponds to the scale factor around $\tau=2$.
The above means that $\tilde{a}(1) \sim 9.5 \times 10^8 \tilde{a}_\text{E}$ at the beginning.
From the 4 dimensional perspective, the scale factor rapidly shrinks from $9.5 \times 10^8 \tilde{a}_\text{E}$ to $\tilde{a}_\text{E}$ 
during ``inflation", and after that it expands like $\tilde{a} \sim \tau^{\frac{9-\sqrt{21}}{20}}$.
The 4 dimensional perspective is appropriate after the inflation where the internal size is negligibly small.

Let us consider the effective action in 4 dimensions.
By using the metric (\ref{eq:met4d}), the dimensional reduction of the action (\ref{eq:W4}) is written as
\begin{alignat}{3}
  S_4 &= \frac{V_7}{2 \kappa_{11}^2} \int d^4x \; \sqrt{- g_\text{E}} 
  \big( R_\text{E} - \tfrac{63}{2} \partial_\alpha \rho \partial^\alpha \rho + \Gamma Z \big), \label{eq:W4-2}
  \\[0.2cm]
  Z &\equiv 24 \big( W_{abcd} W^{abcd} W_{efgh} W^{efgh} - 64 W_{abcd} W^{aefg} W^{bcdh} W_{efgh} \notag
  \\[-0.1cm]
  &\qquad
  + 2 W_{abcd} W^{abef} W^{cdgh} W_{efgh} + 16 W_{acbd} W^{aebf} W^{cgdh} W_{egfh} \notag
  \\
  &\qquad
  - 16 W_{abcd} W^{aefg} W^b{}_{ef}{}^h W^{cd}{}_{gh} - 16 W_{abcd} W^{aefg} W^b{}_{fe}{}^h W^{cd}{}_{gh} \big), \notag
\end{alignat}
where $V_7$ represents the volume of the internal space, and components of the Weyl tensor with spacetime indices are given by
\begin{alignat}{3}
  W_{\alpha\mu\beta\nu} 
  &= e^{-7\rho} \big\{ R^\text{E}_{\alpha\mu\beta\nu} 
  - \tfrac{1}{9} \big( g^\text{E}_{\alpha\beta} R^\text{E}_{\mu\nu} - g^\text{E}_{\alpha\nu} R^\text{E}_{\mu\beta} 
  - g^\text{E}_{\mu\beta} R^\text{E}_{\alpha\nu} + g^\text{E}_{\mu\nu} R^\text{E}_{\alpha\beta} \big) \notag
  \\
  &\quad\,
  + \tfrac{1}{90} (g^\text{E}_{\alpha\beta} g^\text{E}_{\mu\nu} - g^\text{E}_{\alpha\nu} g^\text{E}_{\mu\beta}) R_\text{E} 
  + \tfrac{7}{2} (g^\text{E}_{\alpha\beta} D^\text{E}_\nu \partial_\mu \rho - g^\text{E}_{\alpha\nu} D^\text{E}_\beta \partial_\mu \rho \notag
  \\
  &\quad\,
  - g^\text{E}_{\mu\beta} D^\text{E}_\nu \partial_\alpha \rho + g^\text{E}_{\mu\nu} D^\text{E}_\beta \partial_\alpha \rho )
  - \tfrac{7}{10} (g^\text{E}_{\alpha\beta} g^\text{E}_{\mu\nu} - g^\text{E}_{\alpha\nu} g^\text{E}_{\mu\beta}) D^\text{E}_\gamma \partial^\gamma \rho \notag
  \\
  &\quad\,
  + \tfrac{63}{4} ( g^\text{E}_{\alpha\beta} \partial_\mu \rho \partial_\nu \rho - g^\text{E}_{\alpha\nu} \partial_\mu \rho \partial_\beta \rho
  - g^\text{E}_{\mu\beta} \partial_\alpha \rho \partial_\nu \rho + g^\text{E}_{\mu\nu} \partial_\alpha \rho \partial_\beta \rho ) \notag
  \\
  &\quad\,
  - \tfrac{63}{5} (g^\text{E}_{\alpha\beta} g^\text{E}_{\mu\nu} - g^\text{E}_{\alpha\nu} g^\text{E}_{\mu\beta}) 
  \partial_\gamma \rho \partial^\gamma \rho \big\}, \label{eq:4dWeyl}
  \\
  W_{m\mu n\nu} &= e^{2\rho} \delta_{mn} \big\{ - \tfrac{1}{9} R^\text{E}_{\mu\nu} + \tfrac{1}{90} g^\text{E}_{\mu\nu}R_\text{E}
  - D^\text{E}_\nu \partial_\mu \rho - \tfrac{1}{5} g^\text{E}_{\mu\nu} D^\text{E}_\alpha \partial^\alpha \rho \notag
  \\
  &\quad\,
  - \tfrac{9}{2} \partial_\mu \rho \partial_\nu \rho
  + \tfrac{63}{20} g^\text{E}_{\mu\nu} \partial_\alpha \rho \partial^\alpha \rho \big\}, \notag
  \\
  W_{mpnq} &= e^{11\rho} (\delta_{mn} \delta_{pq} - \delta_{mq} \delta_{np} ) 
  ( \tfrac{1}{90} R_\text{E} - \tfrac{27}{20} \partial_\alpha \rho \partial^\alpha \rho + \tfrac{3}{10} D^\text{E}_\alpha \partial^\alpha \rho ). \notag
\end{alignat}
From this we see that the scalar field $\rho$ couples to the 4 dimensional Riemann tensor $R^\text{E}_{\alpha\mu\beta\nu}$, 
Ricci tensor $R^\text{E}_{\mu\nu}$ and scalar curvature $R_\text{E}$ in a nontrivial way,
and it makes the analysis in 4 dimensions complicate.

Let us give a remark on the derivation of the effective action of perturbations (\ref{eqaction2-0})
from the 4 dimensional action (\ref{eq:W4-2}).
Since the action (\ref{eq:W4-2}) is analyzed perturbatively, it is possible to simplify higher derivatives of perturbative fields
in $Z$ by using equations of motion at $\mathcal{O}(\Gamma^0)$~\cite{Weinberg:2008hq}.
\begin{alignat}{3}
  &R_{\mu\nu} = R^\text{E}_{\mu\nu} + \tfrac{7}{2} g^\text{E}_{\mu\nu} D^\text{E}_\alpha \partial^\alpha \rho
  - \tfrac{63}{2} \partial_\mu \rho \partial_\nu \rho = 0, \notag
  \\
  &R_{mn} = - e^{9\rho} \delta_{mn} D^\text{E}_\alpha \partial^\alpha \rho = 0, \label{eq:4dRicci}
  \\
  &R = e^{7\rho} \big\{ R_\text{E} + 7 D^\text{E}_\alpha \partial^\alpha \rho - \tfrac{63}{2} \partial_\alpha \rho \partial^\alpha \rho \big\} = 0. \notag
\end{alignat}
Note, however, that the above equations of motion should be used up to the linear order of perturbations
although the action (\ref{eq:W4-2}) is expanded up to the second order of the perturbations.
Thus we cannot reduce the Weyl tensor (\ref{eq:4dWeyl}) by simply applying the eqs.~(\ref{eq:4dRicci}) which include 
second order terms. 
In other words, if we want to use the eqs.~(\ref{eq:4dRicci}), we need to cancel the second order perturbations by hand.
(Or we should introduce a second order perturbation field to cancel those terms.)
The equations of motion which are derived from the effective action (\ref{eqaction2-0}) are consistent with
the eqs.~(\ref{eq:P0}) and (\ref{eq:P1}), and we rely on the explicit derivation of the effective action (\ref{eqaction2-0}).

\section{Thoughts on Higher Order Terms} \label{sec:appC}


In this paper we expected that contributions of higher order terms are small compared to the 1-loop order.
However, its reliability depends on the structure of higher order terms in the effective action, which is almost impossible to deal with.
For example, two-loop order effective action consists of a combination of 430 (Weyl)$^7$ terms 
(7 for $W^3W^4$, 17 for $W^2W^5$ and 406 for $W^7$) whose coefficients are unknown.
At best it is possible to consider higher order terms from dimensional analysis. Then $H$ and $G$ are estimated as
\begin{alignat}{3}
  H(\tau) &= \frac{H_\text{I}}{\tau} + \frac{c_h \Gamma H_\text{I}^7}{\tau^7} 
  + \sum_{n=2}^\infty \frac{c_h^{(n)} (\Gamma H_\text{I}^6)^n H_I}{\tau^{6n+1}}, \notag
  \\
  G(\tau) &= \frac{-7 + \sqrt{21}}{14} \frac{H_\text{I}}{\tau} 
  + \frac{c_g \Gamma H_\text{I}^7}{\tau^7}
  + \sum_{n=2}^\infty \frac{c_g^{(n)} (\Gamma H_\text{I}^6)^n H_I}{\tau^{6n+1}}. \label{eq:HGsol-2}
\end{alignat}
Here $\tau$ is given by the eq.~(\ref{eq:tau}), and $c_h$, $c_g$ are given by the eq.~(\ref{eq:chcg}).
Unfortunately, coefficients $c_h^{(n)}$ and $c_g^{(n)}$ are unknown.
As adopted in this paper, if we assume $c_h^{(n)} \sim c_h$, $c_g^{(n)} \sim c_g$ and $\Gamma H_I^6 \ll 1$,
the above expansion is reliable up to the 1-loop order. (Of course, it is also reliable if $\tau$ goes to large, $2 \lesssim \tau$ for instance.)
As a different possibility, the above expansion will also be reliable 
in the case of $c_h^{(n)} \sim (c_h)^n$ and $c_g^{(n)} \sim (c_g)^n$ if we choose $c_h \Gamma H_I^6 \lesssim 1$.
In this appendix, we will examine such possibility.

It is easy to integrate the eq.~(\ref{eq:HGsol-2}). Then $\log a(\tau)$ and $\log b(\tau)$ are solved as
\begin{alignat}{3}
  \log \frac{a}{a_\text{E}} &= \frac{1+\sqrt{21}}{10} \log \tau 
  - \frac{1+\sqrt{21}}{60} \frac{c_h \Gamma H_\text{I}^6}{\tau^6}
  - \frac{1+\sqrt{21}}{60} \sum_{n=2}^\infty \frac{c_h^{(n)} (\Gamma H_\text{I}^6)^n}{n \tau^{6n}}, \notag
  \\
  \log \frac{b}{b_\text{E}} &= - \frac{3\sqrt{21}-7}{70} \log \tau 
  - \frac{1+\sqrt{21}}{60} \frac{c_g \Gamma H_\text{I}^6}{\tau^6}
  - \frac{1+\sqrt{21}}{60} \sum_{n=2}^\infty \frac{c_g^{(n)} (\Gamma H_\text{I}^6)^n}{n \tau^{6n}}. \label{eq:abQc-2}
\end{alignat}
$a_\text{E}$ and $b_\text{E}$ are integral constants, which correspond to scale factors around $\tau=2$.
The behaviors of $a(\tau)$ and $b(\tau)$ depend on the coefficients $c_h^{(n)}$ and $c_g^{(n)}$,
so we will assume those forms below.

The particle horizon $\int \frac{dt}{a(t)}$ during the inflationary era is almost equal to that after the radiation dominated era.
From this, we obtain
\begin{alignat}{3}
  \int_1^2 \frac{a_\text{E}}{a(\tau)} d\tau 
  &= \int_1^2 d\tau \tau^{-\frac{1+\sqrt{21}}{10}}
  \exp \Big[ \frac{1+\sqrt{21}}{60} \Big\{ \frac{c_h \Gamma H_\text{I}^6}{\tau^6} 
  + \sum_{n=2}^\infty \frac{c_h^{(n)} (\Gamma H_\text{I}^6)^n}{n \tau^{6n}} \Big\} \Big] \label{eq:ph1-2}
  \\
  &\sim \frac{9+\sqrt{21}}{6} 2^{\frac{9-\sqrt{21}}{10}} e^{\frac{\sqrt{21}-3}{2} N_\text{e}}. \notag
\end{alignat}
The second line is obtained by using the eq.~(\ref{eq:ph2}).
This gives a relation between $c_h \Gamma H_\text{I}^6$ and $N_\text{e}$ if we know $c_h^{(n)}$.
If we assume $c_h^{(n)} = \frac{(c_h)^n}{n!}$, the second line in the above converges for any $c_h \Gamma H_\text{I}^6$,
and we obtain $c_h \Gamma H_\text{I}^6 \sim 8.5$ for $N_\text{e} = 69$.
On the other hand, if we assume $c_h^{(n)} = (c_h)^n$, the second line in the above converges for $c_h \Gamma H_\text{I}^6 < 1$,
but we obtain $c_h \Gamma H_\text{I}^6 \sim 0.99$ for $N_\text{e} = -1.6$.
Therefore the inflation does not occur in this case.

\begin{figure}[htb]
 \centering
 \begin{picture}(240,185)
 \put(273,153){$a_\text{E} H_\text{I} \eta$}
 \put(-35,0){$\log \frac{\mathcal{P}}{A}$}
 \includegraphics[keepaspectratio,scale=0.75]{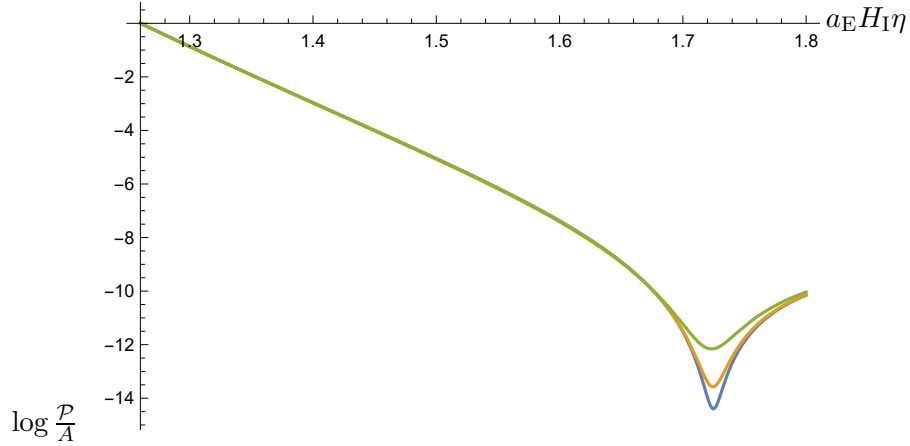}
 \end{picture}
 \caption{Plot of the power spectrum with $c_h \Gamma H_\text{I}^6 = 8.5$.
 $\log \frac{k}{a_\text{E} H_\text{I}}=-30(\text{blue})$, $-20(\text{orange})$, $-10(\text{green})$.}
 \label{fig:PS3}
\end{figure}

\begin{figure}[htb]
 \centering
 \begin{picture}(240,185)
 \put(-45,165){$\log \frac{k}{a_\text{E} H_\text{I}}$}
 \put(273,0){$\log \frac{\mathcal{P}}{A}$}
 \includegraphics[keepaspectratio,scale=0.75]{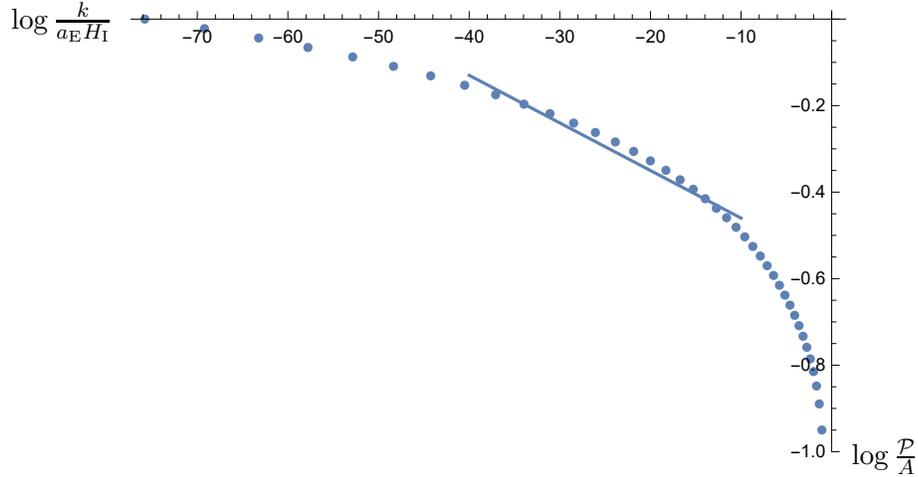}
 \end{picture}
 \caption{Plot of the power spectrum at $k=aH$ with $c_h \Gamma H_\text{I}^6 = 8.5$. The line is given by 
 $\log \frac{\mathcal{P}}{A} = - 0.011 \log \frac{k}{a_\text{E} H_\text{I}} - 0.57$.}
 \label{fig:PS4}
\end{figure}

Let us assume $c_h^{(n)} = \frac{(c_h)^n}{n!}$.
Then the behaviors of the power spectrum (\ref{eq:PS}) with $c_h \Gamma H_\text{I}^6 = 8.5$ and $\log \frac{k}{a_\text{E} H_\text{I}}=-30$, $-20$, $-10$
are plotted in fig.~\ref{fig:PS3}. (We take into account contributions of $2 \leq n$ for the evaluation of $\log (aH)$, 
but neglect those for the calculation of the power spectrum.)
The shapes of the plots change around the end of the inflation, $a_\text{E} H_\text{I} \eta \sim 1.72$, but are similar elsewhere.
The behavior of the power spectrum at the horizon crossing $k=aH$ with $c_h \Gamma H_\text{I}^6 = 8.5$ is shown in fig.~\ref{fig:PS4}.
If we fit the data between $-38 < \log \frac{k}{a_\text{E} H_\text{I}} < -10$ in fig.~\ref{fig:PS4}, 
it is possible to draw a line $\log \frac{\mathcal{P}}{A} = - 0.012 \log \frac{k}{a_\text{E} H_\text{I}} -0.57$, and
the spectral index is estimated as $n_s = 0.99$. (However, contributions from $2 \leq n$ are unknown.)

\newpage


\begin{thebibliography}{99} 



\bibitem{Hiraga:2018kpb}
  K.~Hiraga and Y.~Hyakutake,
  ``Inflationary Cosmology via Quantum Corrections in M-theory'',
  PTEP {\bf 2018} (2018) no.11,  113B03.
  
\bibitem{Ade:2015lrj}
  P.~A.~R.~Ade {\it et al.} [Planck Collaboration],
  ``Planck 2015 results. XX. Constraints on inflation'',
  Astron.\ Astrophys.\  {\bf 594} (2016) A20.

\bibitem{Ade:2015tva}
  P.~A.~R.~Ade {\it et al.} [BICEP2 and Planck Collaborations],
  ``Joint Analysis of BICEP2/Keck Array and Planck Data'',
  Phys.\ Rev.\ Lett.\  {\bf 114} (2015) 101301.

\bibitem{Array:2015xqh}
  P.~A.~R.~Ade {\it et al.} [BICEP2 and Keck Array Collaborations],
  ``Improved Constraints on Cosmology and Foregrounds from BICEP2 and Keck Array Cosmic Microwave Background Data with Inclusion of 95 GHz Band'',
  Phys.\ Rev.\ Lett.\  {\bf 116} (2016) 031302.
  
\bibitem{Starobinsky:1980te}
  A.~A.~Starobinsky,
  ``A New Type of Isotropic Cosmological Models Without Singularity'',
  Phys.\ Lett.\  {\bf 91B} (1980) 99.

\bibitem{Guth:1980zm} 
  A.~H.~Guth,
  ``The Inflationary Universe: A Possible Solution to the Horizon and Flatness Problems'',
  Phys.\ Rev.\ D {\bf 23}, 347 (1981).

\bibitem{Kazanas:1980tx} 
  D.~Kazanas,
  ``Dynamics of the Universe and Spontaneous Symmetry Breaking'',
  Astrophys.\ J.\  {\bf 241}, L59 (1980).

\bibitem{Sato:1980yn} 
  K.~Sato,
  ``First Order Phase Transition of a Vacuum and Expansion of the Universe'',
  Mon.\ Not.\ Roy.\ Astron.\ Soc.\  {\bf 195}, 467 (1981).

\bibitem{Linde:1983gd} 
  A.~D.~Linde,
  ``Chaotic Inflation'',
  Phys.\ Lett.\  {\bf 129B}, 177 (1983).

\bibitem{Freese:1990rb}
  K.~Freese, J.~A.~Frieman and A.~V.~Olinto,
  ``Natural inflation with pseudo - Nambu-Goldstone bosons'',
  Phys.\ Rev.\ Lett.\  {\bf 65} (1990) 3233.

\bibitem{Linde:1993cn} 
  A.~D.~Linde,
  ``Hybrid inflation'',
  Phys.\ Rev.\ D {\bf 49}, 748 (1994)

\bibitem{Boubekeur:2005zm}
  L.~Boubekeur and D.~H.~Lyth,
  ``Hilltop inflation'',
  JCAP {\bf 0507} (2005) 010.

\bibitem{Bezrukov:2007ep}
  F.~L.~Bezrukov and M.~Shaposhnikov,
  ``The Standard Model Higgs boson as the inflaton'',
  Phys.\ Lett.\ B {\bf 659} (2008) 703.

\bibitem{Kolb:1990vq}
  E.~W.~Kolb and M.~S.~Turner,
  ``The Early Universe'',
  Front.\ Phys.\  {\bf 69} (1990) 1.

\bibitem{Liddle:2000cg}
  A.~R.~Liddle and D.~H.~Lyth,
  ``Cosmological inflation and large scale structure'',
  Cambridge, UK: Univ. Pr. (2000).

\bibitem{Dodelson:2003ft}
  S.~Dodelson,
  ``Modern Cosmology'',
  Amsterdam, Netherlands: Academic Pr. (2003).

\bibitem{Weinberg:2008zzc}
  S.~Weinberg,
  ``Cosmology'',
  Oxford, UK: Oxford Univ. Pr. (2008).

\bibitem{Linde:2014nna} 
  A.~Linde,
  ``Inflationary Cosmology after Planck 2013'',
  arXiv:1402.0526 [hep-th].

\bibitem{Dvali:1998pa} 
  G.~R.~Dvali and S.~H.~H.~Tye,
  ``Brane inflation'',
  Phys.\ Lett.\ B {\bf 450}, 72 (1999).

\bibitem{Dvali:2001fw}
  G.~R.~Dvali, Q.~Shafi and S.~Solganik,
  ``D-brane inflation'',
  hep-th/0105203.

\bibitem{GarciaBellido:2001ky}
  J.~Garcia-Bellido, R.~Rabadan and F.~Zamora,
  ``Inflationary scenarios from branes at angles'',
  JHEP {\bf 0201} (2002) 036.

\bibitem{Kachru:2003sx}
  S.~Kachru, R.~Kallosh, A.~D.~Linde, J.~M.~Maldacena, L.~P.~McAllister and S.~P.~Trivedi,
  ``Towards inflation in string theory'',
  JCAP {\bf 0310} (2003) 013.

\bibitem{Silverstein:2008sg}
  E.~Silverstein and A.~Westphal,
  ``Monodromy in the CMB: Gravity Waves and String Inflation'',
  Phys.\ Rev.\ D {\bf 78} (2008) 106003.

\bibitem{McAllister:2008hb}
  L.~McAllister, E.~Silverstein and A.~Westphal,
  ``Gravity Waves and Linear Inflation from Axion Monodromy'',
  Phys.\ Rev.\ D {\bf 82} (2010) 046003.

\bibitem{Baumann:2014nda}
  D.~Baumann and L.~McAllister,
  ``Inflation and String Theory'', Cambridge University Press, 
  arXiv:1404.2601 [hep-th].

\bibitem{Hwang:1996xh}
  J.~c.~Hwang and H.~Noh,
  ``Cosmological perturbations in generalized gravity theories'',
  Phys.\ Rev.\ D {\bf 54} (1996) 1460.

\bibitem{DeFelice:2009ak}
  A.~De Felice and T.~Suyama,
  ``Vacuum structure for scalar cosmological perturbations in Modified Gravity Models'',
  JCAP {\bf 0906} (2009) 034.

\bibitem{DeFelice:2010aj} 
  A.~De Felice and S.~Tsujikawa,
  ``f(R) theories'',
  Living Rev.\ Rel.\  {\bf 13}, 3 (2010).

\bibitem{Sebastiani:2016ras}
  L.~Sebastiani, S.~Vagnozzi and R.~Myrzakulov,
  ``Mimetic gravity: a review of recent developments and applications to cosmology and astrophysics'',
  Adv.\ High Energy Phys.\  {\bf 2017} (2017) 3156915.

\bibitem{Nojiri:2017ncd} 
  S.~Nojiri, S.~D.~Odintsov and V.~K.~Oikonomou,
  ``Modified Gravity Theories on a Nutshell: Inflation, Bounce and Late-time Evolution'',
  Phys.\ Rept.\  {\bf 692}, 1 (2017).

\bibitem{Gross:1986iv}
  D.~J.~Gross and E.~Witten,
  ``Superstring Modifications of Einstein's Equations'',
  Nucl.\ Phys.\ B {\bf 277} (1986) 1.

\bibitem{Gross:1986mw}
  D.~J.~Gross and J.~H.~Sloan,
  ``The Quartic Effective Action for the Heterotic String'',
  Nucl.\ Phys.\ B {\bf 291} (1987) 41.

\bibitem{Grisaru:1986px}
  M.~T.~Grisaru, A.~E.~M.~van de Ven and D.~Zanon,
  ``Four Loop beta Function for the N=1 and N=2 Supersymmetric Nonlinear Sigma Model in Two-Dimensions'',
  Phys.\ Lett.\ B {\bf 173} (1986) 423.

\bibitem{Grisaru:1986vi}
  M.~T.~Grisaru and D.~Zanon,
  ``$\sigma$ Model Superstring Corrections to the Einstein-hilbert Action'',
  Phys.\ Lett.\ B {\bf 177} (1986) 347.

\bibitem{Tseytlin:2000sf}
  A.~A.~Tseytlin,
  ``$R^4$ terms in 11 dimensions and conformal anomaly of (2,0) theory'',
  Nucl.\ Phys.\ B {\bf 584} (2000) 233.

\bibitem{Becker:2001pm} 
  K.~Becker and M.~Becker,
  ``Supersymmetry breaking, M theory and fluxes'',
  JHEP {\bf 0107}, 038 (2001) .

\bibitem{Ishihara:1986if}
  H.~Ishihara,
  ``Cosmological Solutions of the Extended Einstein Gravity With the Gauss-Bonnet Term'',
  Phys.\ Lett.\ B {\bf 179} (1986) 217.

\bibitem{Ohta:2004wk} 
  N.~Ohta,
  ``Accelerating cosmologies and inflation from M/superstring theories'',
  Int.\ J.\ Mod.\ Phys.\ A {\bf 20}, 1 (2005).

\bibitem{Maeda:2004vm}
  K.~i.~Maeda and N.~Ohta,
  ``Inflation from M-theory with fourth-order corrections and large extra dimensions'',
  Phys.\ Lett.\ B {\bf 597} (2004) 400.

\bibitem{Maeda:2004hu}
  K.~i.~Maeda and N.~Ohta,
  ``Inflation from superstring /M theory compactification with higher order corrections. I.'',
  Phys.\ Rev.\ D {\bf 71} (2005) 063520.

\bibitem{Akune:2006dg}
  K.~Akune, K.~i.~Maeda and N.~Ohta,
  ``Inflation from superstring/M-theory compactification with higher order corrections. II. Case of quartic Weyl terms'',
  Phys.\ Rev.\ D {\bf 73} (2006) 103506.

\bibitem{Gibbons:1984kp}
  G.~W.~Gibbons,
  ``Aspects Of Supergravity Theories'',
  Print-85-0061 (CAMBRIDGE).

\bibitem{Maldacena:2000mw}
  J.~M.~Maldacena and C.~Nunez,
  ``Supergravity description of field theories on curved manifolds and a no go theorem'',
  Int.\ J.\ Mod.\ Phys.\ A {\bf 16} (2001) 822.

\bibitem{Gibbons:2003gb}
  G.~W.~Gibbons,
  ``Thoughts on tachyon cosmology'',
  Class.\ Quant.\ Grav.\  {\bf 20} (2003) S321.

\bibitem{Obied:2018sgi}
  G.~Obied, H.~Ooguri, L.~Spodyneiko and C.~Vafa,
  ``De Sitter Space and the Swampland'',
  arXiv:1806.08362 [hep-th].

\bibitem{Hyakutake:2013vwa}
  Y.~Hyakutake,
  ``Quantum near-horizon geometry of a black 0-brane'',
  PTEP {\bf 2014} (2014) 033B04.
  
\bibitem{Mathematicacodes}
  Mathematica codes are located at http://yoshi.sci.ibaraki.ac.jp/arXiv20191025.html

\bibitem{Brandhuber:2001yi} 
  A.~Brandhuber, J.~Gomis, S.~S.~Gubser and S.~Gukov,
  ``Gauge theory at large N and new G(2) holonomy metrics,''
  Nucl.\ Phys.\ B {\bf 611}, 179 (2001)

\bibitem{Brandle:2000qp}
  M.~Brandle, A.~Lukas and B.~A.~Ovrut,
  ``Heterotic M theory cosmology in four-dimensions and five-dimensions'',
  Phys.\ Rev.\ D {\bf 63} (2001) 026003.

\bibitem{Antoniadis:2016avv} 
  I.~Antoniadis and S.~Cotsakis,
  ``Infinity in string cosmology: A review through open problems'',
  Int.\ J.\ Mod.\ Phys.\ D {\bf 26}, no. 04, 1730009 (2016).

\bibitem{Nojiri:2003rz} 
  S.~Nojiri and S.~D.~Odintsov,
  ``Where new gravitational physics comes from: M-Theory?'',
  Phys.\ Lett.\ B {\bf 576}, 5 (2003).

\bibitem{Carroll:2003wy} 
  S.~M.~Carroll, V.~Duvvuri, M.~Trodden and M.~S.~Turner,
  ``Is cosmic speed - up due to new gravitational physics?'',
  Phys.\ Rev.\ D {\bf 70}, 043528 (2004).

\bibitem{Myrzakulov:2015qaa}
  R.~Myrzakulov, L.~Sebastiani and S.~Vagnozzi,
  ``Inflation in $f(R,\phi )$ -theories and mimetic gravity scenario'',
  Eur.\ Phys.\ J.\ C {\bf 75} (2015) 444.

\bibitem{Weinberg:2008hq}
S.~Weinberg,
``Effective Field Theory for Inflation'',
Phys. Rev. D \textbf{77} (2008), 123541.
  

\end{thebibliography}
\end{document}